\documentclass[%
 amsmath,amssymb,
 aps,pre,
 showpacs,
 superscriptaddress,
twocolumn
]{revtex4-2}

\usepackage{graphicx}
\usepackage{dcolumn}
\usepackage{bm}
\usepackage{xcolor}

\usepackage{hyperref}

\newcommand{\red}[1]{\textcolor{black}{#1}}

\newcommand{\out}[1]{ }

\graphicspath{{PRE-NRCH/}}

\begin{document}

\preprint{APS/123-QED}

\title{
Entropy production in the nonreciprocal Cahn-Hilliard model
}

\author{Thomas Suchanek}
 \affiliation{Institut f{\"u}r Theoretische Physik, Universit{\"a}t Leipzig, Postfach 100 920, D-04009 Leipzig, Germany}
\author{Klaus Kroy}
\affiliation{Institut f{\"u}r Theoretische Physik, Universit{\"a}t Leipzig, Postfach 100 920, D-04009 Leipzig, Germany}
\author{Sarah A. M. Loos}
 \email{sl2127@cam.ac.uk}
\affiliation{DAMTP, 
Centre for Mathematical Sciences, 
University of Cambridge, Wilberforce Road, Cambridge CB3 0WA, United Kingdom
}%

\date{\today}

\begin{abstract}
We study the nonreciprocal Cahn-Hilliard model with thermal noise as a prototypical example of a generic class of non-Hermitian stochastic field theories, analyzed in two companion papers [Suchanek, Kroy, Loos, ArXiv:2303.16701 (2023); Suchanek, Kroy, Loos, ArXiv:2305.05633 (2023)]. Due to the nonreciprocal coupling between two field components, the model is inherently out of equilibrium and can be regarded as an active field theory. Beyond the conventional homogeneous and static-demixed phases, it exhibits a traveling-wave phase, which can be entered via either an oscillatory instability or a critical exceptional point. 
By means of a Fourier decomposition of the entropy production rate, we quantify the associated scale-resolved time-reversal symmetry breaking, in all phases and across the transitions, in the low-noise regime. Our perturbative calculation reveals its dependence on the strength of the nonreciprocal coupling. Surging entropy production near the static-dynamic transitions can be attributed to entropy-generating fluctuations in the longest wavelength mode and heralds the emerging traveling wave. Its translational dynamics can be mapped on the dissipative ballistic motion of an active (quasi)particle.
\end{abstract}

\maketitle


\section{Introduction}

A predator chasing fleeing prey~\cite{Wangersky78} or the vision-cones restricting interactions~\cite{barberis2016large,loos2022long,dadhichi2020nonmutual} between herding animals are vivid examples of nonreciprocal interactions, ubiquitous in nonequilibrium many-body systems. Likewise, the effective interactions between colloids in flows~\cite{Hayashi_2006}, active particles~\cite{Liebchen_2022,lavergne2019group}, or in mixtures of active and passive particles \cite{Wittkowski_2017,Mayank17,Mandal22} break the action-reaction principle on the microscale.
Nonreciprocal couplings between different order parameter fields (such as density and polarization fields) also commonly occur in active field theories~\cite{Menzel13,Ghosh2021,Fruchart2021,demarchi2023enzyme}. More generally, recent research has made clear that the concept of nonreciprocal dynamics provides a unifying perspective on a wide range of nonequilibrium phenomena, including waves on membranes~\cite{Ghosh2021}, oscillatory patterns in (binary) fluids~\cite{Mandal22,knevzevic2022,Ses_Sansa_2022,demarchi2023enzyme}, or odd elasticity in soft crystals~\cite{braverman2021topological,Poncet22}. 
In field-theoretical models, nonreciprocal coupling between field components may entail the occurrence of parity-time ($\mathcal{PT}$)-symmetry breaking transitions where dissipative order-parameter patterns emerge, \red{most commonly via oscillatory instabilities \red{(or Hopf-bifurcations)}~\cite{Cross93,cross_greenside_2009} and the the recently uncovered critical exceptional points (CEPs)~\cite{el2018non,krasnok2021paritytime,Fruchart2021},} all of which can be understood from the viewpoint of non-Hermitian dynamics~\cite{Fruchart2021,hanai2020critical}.
\red{Both transition scenario can be characterized by the properties of the \red{critical}
modes 
\red{driving} the transition. For oscillatory instabilities, the eigenvalue of the critical mode is complex and becomes purely imaginary at the transition. At CEPs, a vanishing real eigenvalue is accompanied by the coalescence of unstable modes.} 

The nonreciprocal Cahn-Hilliard model has been recognized as prototypical in this respect~\cite{You_2020,Saha_2020,frohoff2023nonreciprocal}. It exhibits a traveling-wave phase with two associated
$\mathcal{PT}$-symmetry breaking transitions \red{[see Fig.~\ref{fig:pd}(d)]}, one via a line of \red{critical} exceptional points; yet it is simple enough to be studied analytically. 
\red{The traveling wave phase is a paradigmatic example of how dynamical phases of traveling patterns in non-Hermitian field models are always characterized by a coupling of the broken parity of a spatial pattern and the sense of motion, and thus by $\mathcal{PT}$ symmetry breaking.}
The Cahn-Hilliard model consists of two field components, each evolving according to a gradient dynamics, \red{which can be thought of as the densities of two (de)mixing substances.} In the nonreciprocal version, the interaction between the components is not mutually symmetric, and therefore violates the reciprocity principle that is a corner stone of any equilibrium system. The dynamics of such a nonreciprocal field model can therefore not be understood from the perspective of \red{one} global effective potential (or ``nonequilibrium free energy''), but rather as a ``dynamical frustration'': while field A tries to relax to a more favorable state, it pushes field B into a state of higher local energy; and vice versa~\cite{hanai2022nonreciprocal}. By analogy with two particles obeying a run-and-catch relation, this also illustrates the basic mechanism that causes the emergence of traveling patterns. On the level of mode dynamics, a traveling wave can be characterized as a ``chase'' along the Goldstone mode~\cite{Fruchart2021}, which itself tends to dynamically restore a spontaneously broken symmetry.

Compared to the phase behavior of nonreciprocal many-body systems, their fluctuations   are, so far, much less understood. 
A particularly intriguing question concerns the time-reversal symmetry breaking (TRSB) of the fluctuations of nonreciprocal systems, which can be quantified by the (``informatic'') entropy production rate $\mathcal{S}$~\cite{Li_2021,Nardini2017}. 
Indeed, from the analogy with the dynamically frustrated run-and-catch dynamics, we may wonder how the fluctuations reflect the emergence of the collective organization and  mesoscale phases. Conversely, since traveling waves break time-reversal symmetry on the mesoscopic and deterministic level, one may wonder whether this is already the case for transient fluctuations.

In two companion papers~\cite{suchanek2023irreversible,suchanek2023connection} we prove such a statement for a broad class of non-Hermitian models, which includes the nonreciprocal Cahn-Hilliard model studied here. Namely, we show that $\mathcal{S}$ exhibits, quite generally, a characteristic signature at continuous static-dynamic phase transitions.  Here, we provide a more explicit and detailed analysis of TRSB for said model, by means of simulations and perturbative analytical calculations. We explicitly consider $\mathcal{S}$ near the two dissimilar types of static-dynamic phase transitions, namely an oscillatory instability and a CEP, in the low noise regime. 
To this end, we first introduce a Fourier decomposition of the entropy production rate $\mathcal{S}$. It enables us to calculate $\mathcal{S}$ numerically for the full parameter range of the phase diagram, and also provides the basis for our perturbative approaches. 
We find that
$\mathcal{S}$  
increases steeply toward the static-dynamic transitions, as predicted by our general theory~\cite{suchanek2023irreversible,suchanek2023connection}. 
Moreover, our perturbative analytical expressions for the Fourier contributions unravel the scale-dependence of the entropy production, in all phases. 
This is crucial for a qualitative understanding of how TRSB arises, and with which physical features it is associated. 
For example, by extracting the precise scaling of $\mathcal{S}$ in the noise intensity, we can determine whether the fluctuations retain their nonequilibrium character in the deterministic (zero-noise) limit.
Furthermore, an important result, which follows from these analytical considerations, is the role of the {translational fluctuations of the demixing profiles, in the static-demixed phase.}
The latter are shown to account for most of the incipient entropy production. Interestingly, as we rigorously show in Ref.~\cite{suchanek2023irreversible},
the dynamics of the associated patterns
can be mapped onto the persistent motion of an active (quasi)particle, 
illuminating its dissipative character.

The Paper is structured as follows.
In Sec.~\ref{sec:model_framework}, we introduce the model and framework, and present the spectral decomposition of the entropy production rate. 
In Sec.~\ref{sec:Results}, we derive approximative solutions for $\mathcal{S}$ in the low noise regime, studying each of the phases and the associated transition scenarios{, separately}. To tackle $\mathcal{S}$ in the static-demixed phase, we need to develop a refined approach and make some further suitable assumptions.
We reinforce our analytical predictions by comparing them with simulation results, throughout.
To improve readability, some of the technical aspects are given in an Appendix.
%


\section{Framework \& Model}\label{sec:model_framework}

\subsection{Noisy nonreciprocal Cahn-Hilliard model}
\begin{figure*}
\includegraphics[width=.99
\textwidth]{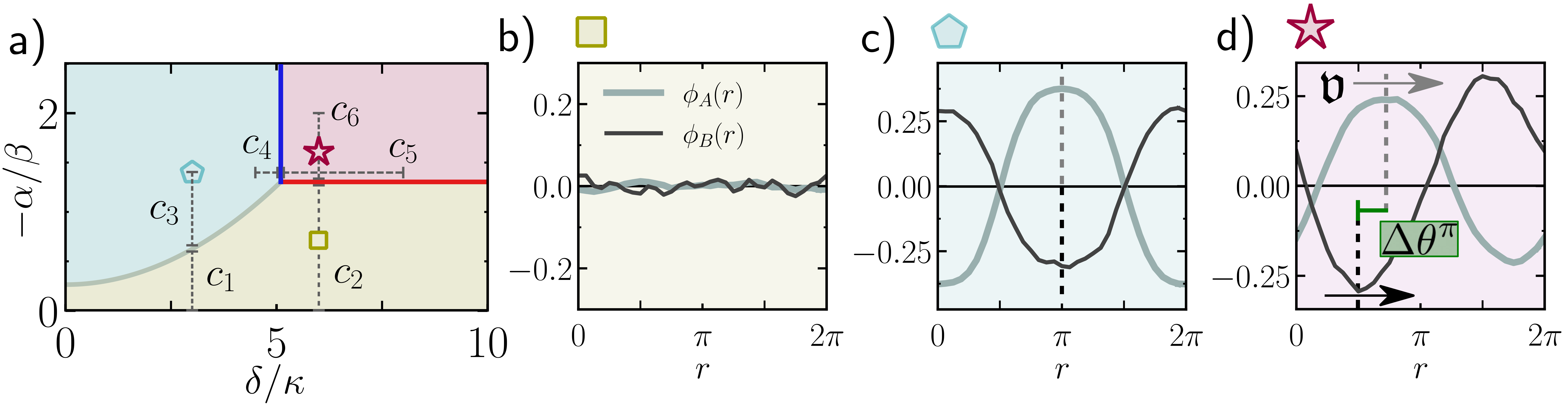}
\caption{\label{fig:pd} 
(a) Phase diagram of the stationary solutions of Eq.~\eqref{equ:1} as established in Ref.~\cite{You_2020} for $\kappa=0.01$, $\gamma=0.015$ and $\beta=0.05$. The mixed phase is marked in yellow, the static-demixed phase in blue and the traveling-wave phase in red;
(b-d) snapshots of the stationary solutions for $\epsilon=2.5\times10^{-5}$. Unlike to the static-demixed phase, the maxima and minima of the profiles of the fields $\phi_A$ (gray) and $\phi_B$ (black) in the traveling-wave phase have a characteristic mean phase shift $\langle \Delta\theta^\pi\rangle>0,$ with $\Delta\theta^\pi\equiv\theta_A-(\theta_B-\pi)$, which is aligned with the direction of propagation. \red{This coupling of the propagation direction to the relative phase shift (which determines the parity of the profile) entails the $\mathcal{PT}$ symmetry breaking in the nonreciprocal Cahn-Hilliard model.}
}
\end{figure*}

We study a stochastic version of the nonreciprocal Cahn-Hilliard model \cite{You_2020,Saha_2020,frohoff2021suppression} of the two-component field ${\phi}(r,t)=(\phi_A,\phi_B)^T$.
The dynamical equations of the two field components read
\begin{equation}
\label{equ:1}
\begin{aligned}
\!\dot \phi_A =& \nabla [ (\alpha\!+\! \phi_A^2 \!-\!\gamma\nabla^2)\nabla \phi_A +(\kappa\!-\!\delta)\nabla\phi_B   + \sqrt{ 2\epsilon} \Lambda_A  ]  \\
\!\dot \phi_B =& \nabla  [\beta\nabla \phi_B +(\kappa+\delta)\nabla\phi_A  
+\sqrt{ 2\epsilon}\Lambda_B  ]\,,
\end{aligned}
\end{equation}
where we use the compact notation, $\dot \phi = \partial_t \phi$. The parameters $\kappa$ and $\delta$ give the strength of the reciprocal (symmetric) and nonreciprocal (asymmetric) coupling, respectively, between the fields. 
We study the dynamics in one spatial dimension on the domain $r \in [0,L]$ with periodic boundary conditions, and focus on the long-time limit $t\to \infty$. 
The noise terms $\sqrt{2\epsilon}{\Lambda}$ are zero-mean, unit-variance, Gaussian white noise fields, constructed such that in the case 
$\delta=0$, the resulting statistical field theory obeys a fluctuation--dissipation relation~\cite{Kubo_1966}, in which $\epsilon$ plays the role of temperature.
\red{The parameters $\alpha$ and $\beta$ and $\gamma$ control the conventional ``demixing'', where $\alpha$ and $\beta$ control the amplitude of the demixed state and $\gamma$ the interface width.} In our numerical examples, we keep $\beta=0.05$ while varying $\alpha$. 
Furthermore, we fix $\kappa=0.01$, corresponding to a weakly repulsive,  reciprocal coupling between the fields, while varying the 
strength of the nonreciprocal coupling
$\delta$.

For some background on the general properties of the model in \red{Eq.~\eqref{equ:1}}, including its phase behavior, the morphology of its stationary states, and coarsening dynamics, we refer the Reader to Refs.~\cite{You_2020,Saha_2020,frohoff2021suppression,frohoff2023non}.

\subsubsection{Phase behavior}
The stationary solutions of Eq.~\eqref{equ:1} in the noise-free limit, which we denote by ${\phi}^*$, 
can be parametrized as
\begin{align}\label{equ:phaseamp}
\phi^*_{A,B}(r,t)=\sum\limits_{k>0}\mathcal{A}^{k,*}_{A,B}\cos\left[q_k r+\theta^{k,*}_{A,B}(t)\right],
\end{align}
by Fourier modes with wavenumbers $q_k\equiv 2\pi k /L$ for odd $k \in \{2n+1:n\in\mathbb{N}_0\}$, \red{amplitudes  $\mathcal{A}^{k,*}_{A,B}$ and phases $\theta^{k,*}_{A,B}(t)$.} It is known that ${\phi}^*$ exhibits three distinct phases as depicted in Fig.~\ref{fig:pd}~\cite{You_2020}:
first a homogeneous, or ``mixed'' phase ($\mathcal{A}^{k,*}_{A,B}=0$) for positive and small negative $\alpha$, which gives way to a second, inhomogeneous or ``demixed'' state upon a decrease of $\alpha$ below $\alpha_c$. For negative or small positive values of the nonreciprocity $\delta$, the demixed state is static with $\mathcal{A}^{k,*}_{A,B}>0$, $\theta^{k,*}_{A,B}(t)=\mathrm{const}$. In contrast, if $\delta$ exceeds the critical threshold $\delta_c\equiv\sqrt{\beta^2+\kappa^2}$, the model admits a third phase with traveling-wave solutions ($\mathcal{A}^{k,*}_{A,B}>0$,   $\dot\theta^{k,*}_{A,B}/q_1=\mathfrak{v}$). Its propagation velocity is $\mathfrak{v} = \pm \sqrt{\delta^2-\delta_c^2}$, and its direction is randomly set by the noise and initial condition~\cite{You_2020}, such that, for an ensemble with random initial conditions and noise, $\big\langle \dot\theta^{k,*}_{A,B}\big\rangle =0$, with $\left\langle .\right\rangle$ denoting the noise average. 
{The direction of the wave propagation is (on average) aligned with a characteristic phase shift $\Delta\theta^{*,\pi} = \theta^*_A-(\theta^*_B-\pi) $, i.e., with the parity of the wave profiles~\cite{suchanek2023irreversible,You_2020}, see Fig.~\ref{fig:pd}(d) for an illustration.}

The transition from the homogeneous to the traveling-wave state occurs through an oscillatory instability [solid red line in Fig.~\ref{fig:pd}(a)]. The transition from the static-demixed state to the traveling-wave state is a line of critical exceptional points \cite{Ashida2020} [solid blue line in Fig.~\ref{fig:pd}(a)]. For an in-depth description of the characteristics of critical exceptional points, in this context, we refer to Refs.~\cite{Fruchart2021,suchanek2023connection}. 
Furthermore, it has been shown that in a fairly large region of the phase diagram above the demixing transitions, the first Fourier mode is dominant in the two demixed states~\cite{You_2020}, i.e., $\mathcal{A}^{1,*}\gg \mathcal{A}^{k,*}$ for all $k>1$.

Our numerical investigations of Eq.~\eqref{equ:1} reveal that in the regime of small $\epsilon>0$ the \red{phase diagram in Fig.~\ref{fig:pd} remains essentially unchanged compared to the zero-noise case. Yet the transition points widen into an extended region} (details about the numerics are given in Sec.~\ref{sec:numerics} and App.~\ref{sec:num}). 
\red{Nevertheless, it is important to take noise into account, because even a small intensity noise brings to light strong fluctuations, which
appear exclusively in the vicinity of the phase transitions --- in analogy to equilibrium critical phenomena. However, in sharp contrast to equilibrium critical phenomena, here these strong fluctuations possess strong time-reversal asymmetry, leading to new phenomena such as (long-lived) transient traveling waves and actively enhanced interfacial dynamics, which substantially alter the appearance of the stationary state, as we demonstrate in Ref.~\cite{suchanek2023irreversible}. In this article, we investigate the entropy production associated with these new noise-induced phenomena in more detail.}

\subsubsection{Nonequilibrium currents and model structure}
The dynamical equations~\eqref{equ:1} can be recast into the 
following form to better reveal their general structure,
\begin{align}\label{equ:mod}
    \dot \phi_{A,B}&=-\nabla\cdot(J^\mathrm{d}_{A,B}+\sqrt{2\epsilon}\Lambda_{A,B}),
    \\
    J^\mathrm{d}_{A,B}&=-\nabla\left( \frac{\delta\mathcal{F}}{\delta\phi_{A,B}}+ \mu^a_{A,B}\right)\,.
    \label{equ:modcurrent}
\end{align}
The deterministic current in \red{Eq.~\eqref{equ:modcurrent}} is composed of two parts. 
Its equilibrium-like contribution derives from a scalar potential
\red{
\begin{align}
\mathcal{F}=\int_V\mathrm{d}r \frac{1}{2}\left[\alpha\phi_A^2+\frac{1}{6}\phi_A^4+\beta\phi_B^2+\gamma(\nabla\phi_A)^2 \right],
\end{align}
}corresponding to the free energy of a standard Cahn-Hilliard field $\phi_A$ coupled to a Gaussian field $\phi_B$. Its 
nonequilibrium component is generated by a nonequilibrium chemical potential 
\begin{align}\label{eq:chemicalpotential}
{\mu}^{a}=\delta\, (- \phi_B, \phi_A )^T.
\end{align}
Since it cannot be represented as variation of a scalar potential, ${\mu}^a$ is a source of currents that are not compatible with thermal equilibrium. 
This decomposition unveils that the nonreciprocity of the coupling between $\phi_A$ and $\phi_B$ implies by itself  a deviation from equilibrium-like dynamics. Note that other decompositions can be chosen, but the one in Eqs.~\eqref{equ:mod}, \eqref{equ:modcurrent} clearly separates equilibrium-compatible and inherently nonequilibrium dynamics~\cite{O'Byrne21}. 

\subsection{Measuring irreversibility}
A suitable {measure} of the nonequilibrium character of mesoscopic models 
is provided by the TRSB of their fluctuations \cite{Li_2021,Nardini2017,Seifert2005}.
We quantify the latter for Eq.~\eqref{equ:1}, using the (informatic) entropy production rate for field models and the techniques developed in Refs.~\cite{Li_2021,Nardini2017,suchanek2023connection}. In this context, the entropy production along a given trajectory $\{{\phi}_{t\in [0,T]}\}$
 is defined as the log ratio of the path probabilities~\footnote{To be well-defined, the ratio of path probabilities in Eq.~\eqref{equ:probratio} is to be interpreted in terms of the Onsager–Machlup formalism~\cite{Onsager53}.} for the observation of that trajectory and its time-reversed realization $\{{\phi}^R_{t\in [0,T]}\}$,
\begin{equation}\label{equ:probratio}
     {s}[{\phi};0,T]\equiv\log \frac{\mathbb{P}\left[\{{\phi}_{t\in [0,T]}\}\right]}{\mathbb{P}\left[\{{\phi}^R_{t\in [0,T]}\}\right]}\, .
\end{equation}
The fields are assumed to be position-like and thus even under time reversal. The average entropy production rate 
\begin{align}
{{\mathcal{S}}}= \lim_{h\to 0} \left\langle{s}[{\phi};t,t+h]/{h}\right\rangle
\end{align}
serves as a measure of the average TRSB at time $t$.
By construction, $\mathcal{S}$ is constant in time for any steady state. In particular, $\mathcal{S}=0$ in thermal equilibrium, where ${\mu}^a\!=\!0$.

In the vicinity of the phase transitions, the \red{zero-noise} limit 
\begin{align}
\mathcal{S}^*=\lim_{\epsilon\to 0}\mathcal{S}
\end{align}
is particularly interesting~\cite{suchanek2023irreversible}. While not a natural observable itself,  $\mathcal{S}^*$ is a means to extract quantitative information about the TRSB at {leading order} in $\epsilon$.
In Refs.~\cite{suchanek2023irreversible,suchanek2023connection}, we have shown that for models of \red{the type of Eq.}~\eqref{equ:mod}, $\mathcal{S}^*$ exhibits a characteristic behavior across continuous phase transitions from a static to a dynamical phase. {Specifically, both for oscillatory instabilities and CEPs, we have demonstrated that $\mathcal{S}$ generally surges toward the transition point and scales like the susceptibility.} In the following, we present a detailed analytical and numerical investigation of this phenomenon for the concrete model \red{of Eq.}~\eqref{equ:1}, in the low noise limit.


\subsection{Decomposition of $\mathcal{S}$}

We first derive a decomposition of $\mathcal{S}$ in terms of contributions from different Fourier modes, which forms the basis for our analytical and numerical investigations, and provides insights into TRSB at different scales.

We start from the general expression for models \red{of the type of Eq.}~\eqref{equ:mod}
\begin{align}\label{equ:EPR1}
\mathcal{S}=&\int_V\mathrm{d}r
\frac{\sum_i\left\langle J^\mathrm{d}_i\nabla{\mu}^a_i\right\rangle}{\epsilon}-
\int_V\mathrm{d}r\sum_i\left\langle\frac{\delta}{\delta{\phi_i}}\nabla^2\mu_i^a \right\rangle
    \,,
\end{align}
which we derived in Ref.~\cite{suchanek2023connection}. 
Exploiting that the nonequilibrium chemical potential \red{given in Eq.~\eqref{eq:chemicalpotential}}
is totally antisymmetric in the field components, Eq.~\eqref{equ:EPR1} simplifies to 
\begin{align}
    \mathcal{S}=\int_0^L \!\!\mathrm{d}r \frac{\delta}{\epsilon}\,\Big[\! \left\langle J^\mathrm{d}_A(r) \nabla\phi_B(r)\right\rangle-\left\langle J^\mathrm{d}_B(r) \nabla\phi_A(r)\right\rangle\!\Big].
\end{align}
Further, using that, in the stationary state,
\begin{align}
    0&=\int_0^L\mathrm{d}r \partial_t \langle {\phi_A\phi_B}\rangle
    \nonumber =\int_0^L\mathrm{d}r\big(\langle J^\mathrm{d}_A\nabla\phi_B\rangle+\langle J^\mathrm{d}_B\nabla\phi_A\rangle\big),
\end{align}
we find
\begin{align}\label{equ:EPRcurrent}
    \mathcal{S}=-\epsilon^{-1}2\delta\int_0^L\mathrm{d}r \left\langle J_B^\mathrm{d}\nabla\phi_A \right\rangle
    \,.
\end{align}
Now we transform the last expression to Fourier space, according to 
\begin{align}
    {\phi}^k_{A,B}\equiv L^{-1}\int_{0}^{L}\mathrm{d}r\, \phi_{A,B}(r) e^{-iq_kr}\,.
\end{align}
Plugging in the explicit expression \red{Eq.~\eqref{equ:modcurrent}} for $J^\mathrm{d}_B$, we find the linear decomposition
\begin{align}\label{equ:esum}
\mathcal{S}=\sum_{k=1}^\infty \sigma^k\, ,
\end{align}
with Fourier components
\begin{align}\label{eq:EPRmodes}
    \sigma^k=
    \frac{\delta 4L q_k^2}{\epsilon}
    \left[ \beta\mathrm{Re}\left\langle  {\phi}^k_{A}{\phi}^{-k}_{B}\right\rangle+ (\kappa+\delta) \Big\langle \vert{\phi}_{A}^k\vert^2\Big\rangle\right]
    \,.
\end{align}

Importantly, although the nominal mode contribution $\sigma^k$ to the entropy production rate depends only on the statistics of ${\phi}^k$, 
it cannot strictly be assigned exclusively to the dynamics of ${\phi}^k$. 
Recalling the definition of $\mathcal{S}$ in terms of the path probabilities of the full process ${\phi}$, such an assignment would only be legitimate if the dynamics of the $k$th mode was statistically {independent} of all other modes.
Indeed, in this case, the entropy production rate of the $k$th mode would be a well-defined quantity, and equal to $\sigma^k$ as given in Eq.~\eqref{eq:EPRmodes}. For our model, this is never strictly the case. However, such stochastic independence indeed approximately holds under certain conditions, which we discuss in detail in Sec.~\ref{sec:Results}.

In any case, Eq. \eqref{eq:EPRmodes} provides a representation of $\mathcal{S}$ exclusively in terms of quadratically expectation values of $\phi_{A,B}$. 
The latter can be calculated perturbatively up to linear order in $\epsilon$ to make further analytical progress (see Sec.~\ref{sec:Results}) and to determine $\mathcal{S}^*$ exactly. 
As a bonus, the representation \red{given in Eq.}~\eqref{eq:EPRmodes}
is also convenient for numerical calculations of $\mathcal{S}$ throughout the phase diagram (see App. \ref{sec:eprhom}).
%
\begin{figure}
\includegraphics[width=.45\textwidth]{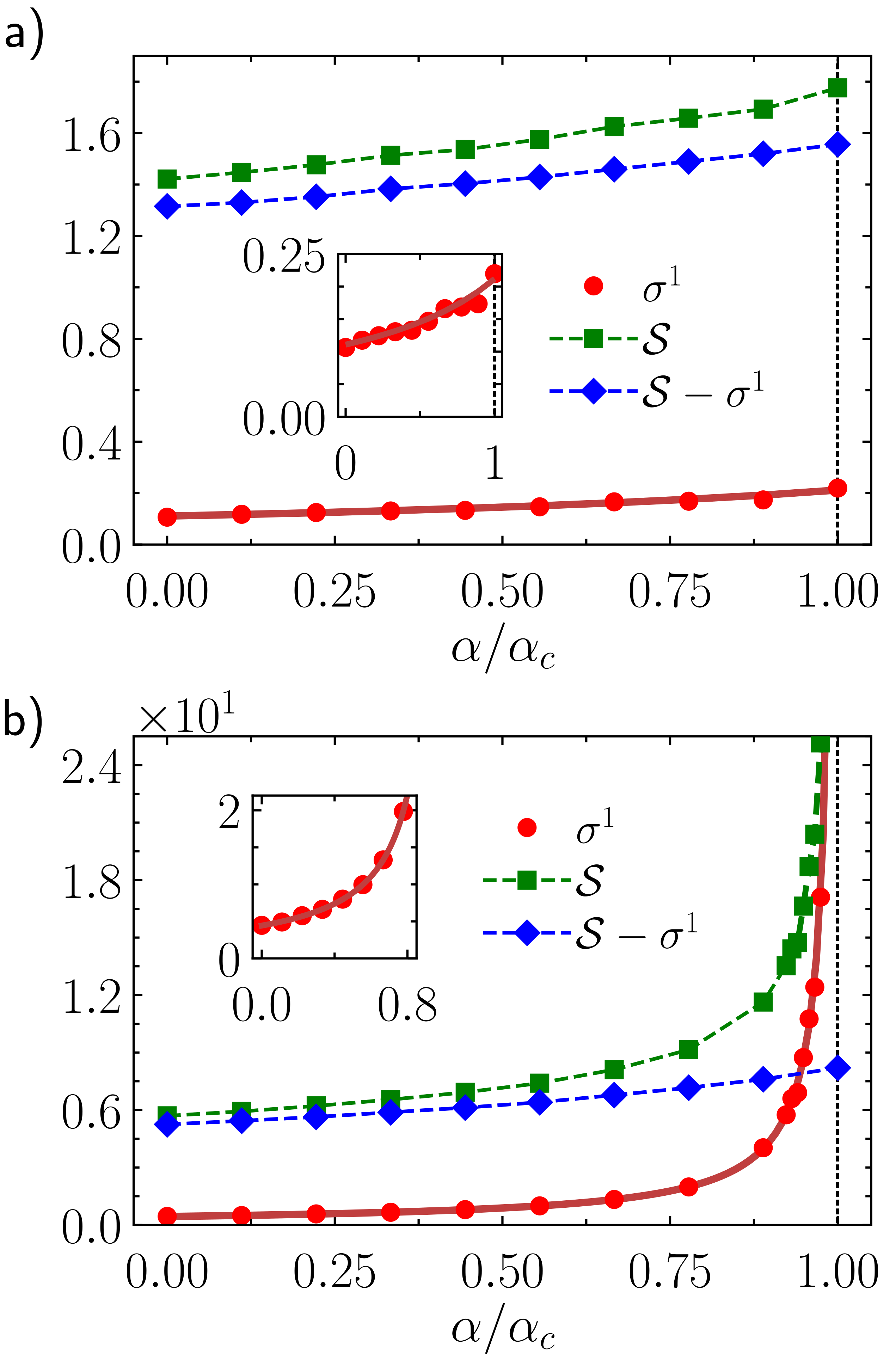}
\caption{\label{fig:hom} 
Entropy production rate \red{$\mathcal{S}$ and the first-mode contribution $\sigma^1$ 
\red{[defined in 
Eqs.~(\ref{equ:esum}, \ref{eq:EPRmodes})]}
} as a function of the ``demixing'' parameter $\alpha$ in the homogeneous phase in the vicinity of the phase boundaries (at $\alpha_c$, indicated by dashed vertical lines), respectively: (a) near the transition to the static-demixed phase [here, $\delta=0.8\delta_c$, path $c_1$ in Fig.~\ref{fig:pd}(a)]; (b) near the traveling-wave phase (here, $\delta\!=1.2\delta_c$, path $c_2$ in Fig.~\ref{fig:pd}).
The analytical prediction \red{of Eq.}~\eqref{res1} is shown by solid red lines. Symbols depict simulation results, dashed lines serve as guides to the eye. Insets depict zooms into a smaller $y-$axis range.
We observe very distinct behavior near the transitions. 
Toward the static-static transition [in~(a)], the entropy production rate only mildly increases but stays regular. Toward the static-dynamic transition [in~(b)], the first-mode contribution $\sigma^1$, and hence also $\mathcal{S}$, increase steeply, and formally diverges in the zero-noise limit, even after UV regularization.
 Other parameters are $\kappa =0.01$, $\gamma =0.015$, $\beta =0.05$, $\epsilon =10^{-10}$.
}
\end{figure}

\subsection{Numerical method and UV cutoff}\label{sec:numerics}

To numerically integrate Eq.~\eqref{equ:1} for small values of $\epsilon$ we used a Euler-Maruyama algorithm with finite difference gradients on a domain of length $L=2\pi$, discretized by $32$ equally spaced mesh points, and time slices $\Delta t= 0.01$. 
We determine the numerical value of ${\phi}^k_{A,B}$ as well as the statistics of $\dot\phi^k_{A,B}$. 
For the mixed phase, we use the latter to numerically evaluate the mode contributions $\sigma^k$ as defined in Eq.~\eqref{eq:EPRmodes}. For the demixed phases, we use an alternative route to calculate $\mathcal{S}$ based on the decomposition of phase and amplitude [as in Eq.~\eqref{equ:phaseamp}] described in App.~\ref{sec:num}. It offers the possibility to separately calculate the contribution due to the fluctuations that lead to displacements of the interfaces between $\phi_A$ and $\phi_B$, which are of special interest~\cite{suchanek2023irreversible}. It also turned out to provide a much faster numerical convergence.

{A general observation from the numerical data is that, in the steady state and low noise regime, only the lowest modes are coupled, in all three phases.} 
Furthermore, consistent with our analytical results presented below in Eq.~\eqref{eq:sigmainf} and App.~\ref{sec:eprhom}, $\sigma^k$ quickly converges to a function 
that depends on $\delta$ and $\gamma$ only, in all phases. 
The underlying mathematical reason is that for higher $k$, the {highest-order differential operator} $\nabla^3$ in Eq.~\eqref{equ:1} completely dominates $\boldsymbol{J}^\mathrm{d}$ (see App.~\ref{sec:eprhom}). 
This also implies that $\mathcal{S}$ is always formally divergent, as a consequence of the usual ``UV catastrophe'', which is cured by introducing a cutoff length scale [the cutoff length scale can be thought of as the maximum resolution with which the process \red{defined in Eq.~\eqref{equ:1}} is observed, or a microscale beyond which the continuous field description breaks down].
Our numerical computations showed that, for the parameter regime we investigated, the choice $k=5$ is appropriate to study the change of the behavior of $\mathcal{S}$. 
Further details are provided in App.~\ref{sec:num}. Moreover, in App.~\ref{Sec:cutoff}, we discuss possibilities to introduce a cutoff length without spoiling the genuinely probabilistic interpretation of $\mathcal{S}$ in terms of path probabilities.

\begin{figure*}
\includegraphics[width=.99\textwidth]{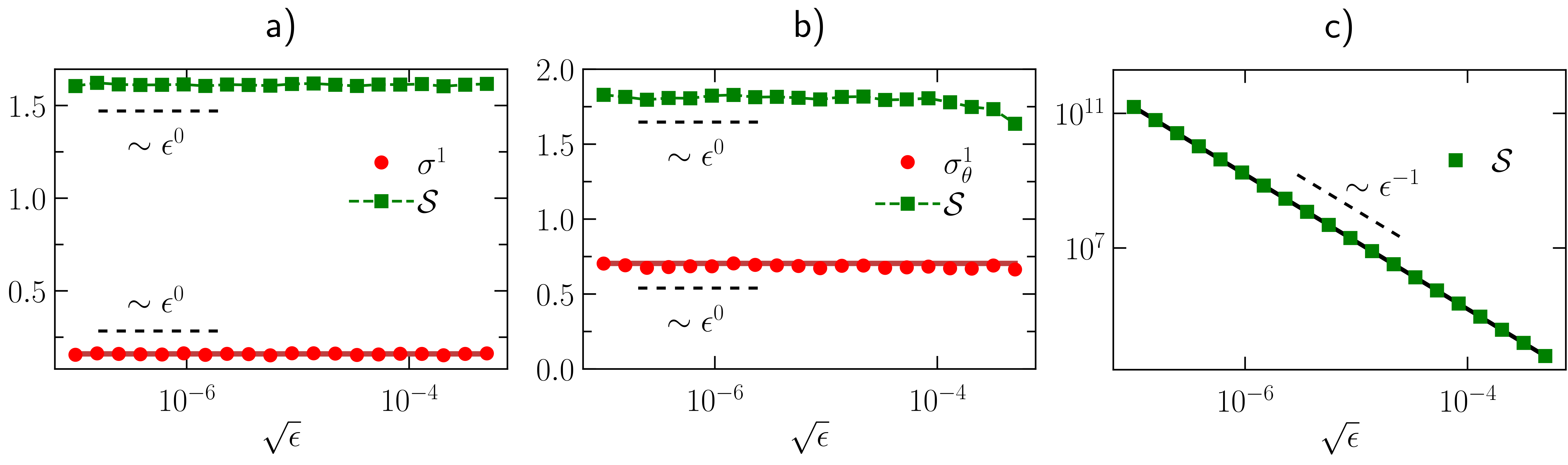}
\caption{\label{fig:scaling} 
Scaling of the entropy production rate \red{$\mathcal{S}$ and the first-mode contribution $\sigma^1$ \red{[defined in 
Eqs.~(\ref{equ:esum}, \ref{eq:EPRmodes}), shown here in panel (a)]}, $\mathcal{S}$ and 
$\sigma_\theta^1$ 
\red{[defined in 
Eqs.~(\ref{EPRnumtotal}, \ref{def:Spnum_k}), and shown in panels (b,c)]}} with the noise intensity $\epsilon$, for $\epsilon\rightarrow0$ within the different phases: (a) Mixed phase ($\delta=0.03, \alpha=-0.02$); (b) static-demixed phase ($\delta=0.045, \alpha=-0.07$); (c)
traveling-wave phase ($\delta=0.06, \alpha=-0.07$). The analytical results for the low noise limits given in Eqs.~\eqref{res1}, \eqref{eq:theta1contribution}, \eqref{eq:EPR-travelingwave} are shown by solid lines.
The numerical results confirm the analytics, including that $\mathcal{S}\sim \epsilon^0$ in the static phases, and $\mathcal{S} \sim \epsilon^{-1}$ in the traveling-wave phase.
Other parameters are $\kappa=0.01, \gamma=0.015, \beta=0.05$.
}
\end{figure*}


\section{Irreversibility in the different phases and across the transitions}\label{sec:Results}

Equipped with the framework described in Sec.~\ref{sec:model_framework}, we now analytically and numerically investigate TRSB in the model \red{given in Eq.~\eqref{equ:1}}.
We study the homogeneous and the inhomogeneous parts of the phase diagram separately.

\subsection{Homogeneous phase}
\label{sec:homentropy}
Consider first the mixed, homogeneous phase, i.e., the subdomain of the phase diagram where ${\phi}^*=0$. Here, we can determine $\mathcal{S}$ up to leading order in $\epsilon$, using the standard approach of the small noise expansion~\cite{gardiner2009}.
Each Fourier mode is expanded as
\begin{align}
{\phi}_{A,B}^k=\sqrt{\epsilon}\,{\psi}_{A,B}^k+\mathcal{O}(\mathcal{\epsilon}). 
\end{align}
Inserting this Ansatz into Eq.~\eqref{equ:1}, leads to an $\epsilon$-independent dynamical equation
\begin{align}\label{equ:eomm}
\dot{{{\psi}}}^k(t)= \mathbb{A}_k{{\psi}}^k(t)+{\xi}^k\,,
\end{align}
for the leading order expansion coefficient, where
\begin{align}
 \mathbb{A}_k=-q_k^2
\begin{pmatrix}
\alpha+\gamma q_k^2  & \kappa-\delta\\
\kappa+\delta & \beta\\
\end{pmatrix} \,
\end{align}
is the dynamical operator obtained by linearizing Eq.~\eqref{equ:1}
and $\left\langle\xi^k_A\xi^l_A\right\rangle=\left\langle\xi^k_B\xi^l_B\right\rangle=2q_ k^2/L \,\delta_{k,-l}$, $\left\langle\xi^k_A\xi^l_B\right\rangle=0$. Thus, in the first-order approximation, the dynamics of the different modes in Fourier space is fully decoupled, and $\sigma^k$ can be interpreted as a measure of TRSB by the $k$-th mode.
The elements of the covariance matrix of ${{\psi}}^k$ (see App.~\ref{sec:eprhom}) are obtained by solving  the Lyapunov equation. Noting that $\left\langle{\phi}^{-k}_i{\phi}^k_j\right\rangle=\epsilon\left\langle{\psi}_i^{-k}{\psi}_j^k\right\rangle+\mathcal{O}(\epsilon^2)$, $i,j\in \{A,B\}$ and plugging the result into Eq.~\eqref{eq:EPRmodes}, we obtain
\begin{align}\label{res1}
   \sigma^k
    =\frac{8\delta^2 \,q_k^2}{\alpha+\beta+\gamma q_k^2}+\mathcal{O}(\epsilon)
    \,.
\end{align}
Equation~\eqref{res1} implies that it is the nonreciprocal coupling between the field components that leads to entropy production, on the scale of each Fourier mode. Specifically, each contribution depends {quadratically} on the degree of nonreciprocity, $\sigma^k\sim \delta^2$, which is indicative of a purely nonreciprocal origin of the entropy production~\footnote{Indeed, the sole presence of a nonreciprocal coupling in a system of two overdamped or underdamped particles would also yield an entropy production rate $\sim \delta^2$~\cite{loos2020irreversibility}.}. 
Accordingly, also $\mathcal{S}\sim \delta^2$.

An important implication of Eq.~\eqref{res1} is the qualitatively different behavior of $\mathcal{S}$ toward the phase boundaries to the static-demixed phase versus the dynamical phase. At the transition to the static-demixed phase, the denominator in Eq.~\eqref{res1} remains finite for all $k$, resulting in finite $\mathcal{S}^*$. 
In contrast, toward the transition to the traveling-wave phase, the denominator only remains finite for $k>1$, but vanishes for $k=1$. Hence, \red{Eq.~\eqref{res1}} implies a divergence of $\mathcal{S}^*$, at the transition, which can be exclusively attributed to the first mode, $\sigma^1$.
The numerical data displayed in Fig.~\ref{fig:hom}(a,b) indeed supports these predictions. 
Equation~\eqref{res1} further predicts that $\mathcal{S}$ scales like $\sim \epsilon^0$, i.e., is independent of the noise intensity at leading order,
in the homogeneous phase. 
The comparison with the numerical data displayed in Fig.~\ref{fig:hom}(a,b) and Fig.~\ref{fig:scaling}(a) shows good agreement.

Furthermore, Eq.~\eqref{res1} captures another notable feature, which is confirmed by the simulation data, namely that, in the limit of small length scales, the entropy production components
approach a finite positive value of 
\begin{align}\label{eq:sigmainf}
\lim_{k\rightarrow \infty}\sigma^k=8\delta^2/\gamma,
\end{align}
which is independent of $\alpha$ and $\beta$. As we explicitly show in App.~\ref{sec:eprhom}, this does not only hold for the homogeneous phase, but Eq.~\eqref{eq:sigmainf} is in fact valid in the whole parameter range of the phase diagram, indicative of a ``UV catastrophe''. 
The corresponding modes can be regarded as spurious degrees of freedom that could be suppressed for the present purpose, thus providing a motivation for the regularization of Eq.~\eqref{equ:esum} through a UV cutoff. 

\subsection{Static-demixed phase}\label{sec:static-demixed}

For the static-demixed phase (where ${\phi}^*\neq 0$), the analytical investigation turns out to be more involved than for the homogeneous phase, and requires additional steps. The first complication lies in the fact that we cannot immediately apply the small noise expansion.
For a given deterministic solution ${\phi}^*$, $\left\langle\vert \phi-\phi^* \vert^2 \right\rangle$ is of order $\epsilon^0$, so that we cannot directly expand $\phi-\phi^*$ in orders of $\epsilon^{1/2}$. The underlying reason is that, in the presence of noise, the position of the interfaces of the demixed state does not remain fixed, but diffuses freely. To resolve this issue, we resort to the amplitude-phase representation
\begin{align}
{\phi}^k_{A,B}=\frac{\mathcal{A}_{A,B}^k}{2} \mathrm{exp}\{i \theta_{A,B}^k\}
\end{align}
and consider fluctuations of $\mathcal{A}^k_{A,B}$ and $\theta^k_{A,B}$, separately, {within a small noise expansion}. This requires the exact dynamical equations for the amplitude an the phase, for which we find
\begin{widetext}
\begin{equation}
\begin{aligned}\label{equ:dynex1}
    \dot{\mathcal{A}}_A^k=&-q_k^2\left[(\alpha + \gamma q_k^2) \mathcal{A}_A^k+(\kappa-\delta)\mathcal{A}_B^k\cos(\theta_A^k-\theta_B^k)+\mathrm{Re}(\mathrm{K}^k)-\frac{4\epsilon}{L\mathcal{A}_A^k}\right]+\xi_{\mathcal{A}_A^k}\\ 
    \dot{\mathcal{A}}_B^k=&-q_k^2\left[\beta \mathcal{A}_B^k+(\kappa+\delta)\mathcal{A}_A^k\cos(\theta_A^k-\theta_B^k)-\frac{4\epsilon}{L\mathcal{A}_A^k}\right]
    +\xi_{\mathcal{A}_B^k}\,,
\end{aligned}
\end{equation}
\end{widetext}
and
\begin{equation}
\begin{aligned}\label{equ:dynex2}
  \!\!  \dot{\theta}_A^k=&\frac{q_k^2}{\mathcal{A}_A^k}\left[\sin(\theta_A^k-\theta_B^k)\mathcal{A}_B^k(\kappa-\delta)-
    \mathrm{Im}(\mathrm{K}^k)\right]+\xi_{\theta_A^k}\\
   \!\! \dot{\theta}_B^k=&-\frac{q_k^2}{\mathcal{A}_B^k}\sin(\theta_A^k-\theta_B^k) {\mathcal{A}_A^k(\kappa+\delta)}+\xi_{\theta_B^k},
\end{aligned}
\end{equation}
with
\begin{align}\label{eq:coupling}
    \mathrm{K}^k= \sum\limits_{l,l'}\frac{q_{l}^2+2q_{l}q_{l'}}{4q_k^2}
    {\mathcal{A}^k_A\mathcal{A}^l}_A\mathcal{A}_A^{k-l'-l}
    e^{i(\theta_A^l+\theta_A^{l'}+\theta_A^{k-l'-l}-{i\theta_A^k})}.
\end{align}
The co-variance matrix of the transformed noise is still diagonal with
\begin{align}
\left\langle\xi_{\mathcal{A}_{A,B}^k},\xi_{\mathcal{A}_{A,B}^l}\right\rangle&=q_k^2\frac{4\epsilon}{L}\delta_{kl} \, ,
\\
\left\langle\xi_{\theta_{A,B}^k},\xi_{\theta_{A,B}^l}\right\rangle&=q_k^2\frac{4\epsilon}{L\mathcal{A}_{A,B}^2}\delta_{kl}\, , \\
%
\left\langle\xi_{\mathcal{A}_{A}^k},\xi_{\mathcal{A}_{B}^l}\right\rangle&=\left\langle\xi_{\theta_{A}^k},\xi_{\theta_{B}^l}\right\rangle=0\,, 
\end{align}
but the transformation has rendered the noise  \textit{multiplicative}. 
From Eqs.~\eqref{equ:dynex1} and \eqref{equ:dynex2}, we clearly see that unlike for ${\phi}^*\neq0$, fluctuations with different wavenumbers in general do not decouple. This is a general feature of any nonlinear field model. 

As in Eq.~\eqref{equ:eomm}, treating the amplitude and phase dynamics within the small noise expansion up to first order,  corresponds to linearizing Eqs.~\eqref{equ:dynex1} and \eqref{equ:dynex2}. As we explicitly show in App.~\ref{app:ap}, the linearized (low-noise) dynamics admits a separation into two {statistically independent} parts: 
{The first part comprises all the amplitude fluctuations $\mathcal{A}_{A,B}^k$ with odd $k\in\{2n+1 : n\in \mathbb{N}_0\}$, as well as field fluctuations $\phi^k_{A,B}(t)$ of even $k\in\{2n : n\in \mathbb{N}_0\}$. This can be interpreted as the collection of fluctuations altering the amplitudes and distorting the demixing profiles of $\phi_A$ and $\phi_B$, at a fixed position.
The second part consists of the $\theta^k_{A,B}(t)$ fluctuations for odd $k\in\{2n+1 : n\in \mathbb{N}_0\}$, which may lead to displacements of the demixing profiles, as well as their dispersion.
Importantly, this also includes relative translations of the profiles of fields A and B creating a characteristic phase shift $\Delta\theta^\pi=\theta_A-(\theta_B-\pi)\neq0$, as well as joint translations. The latter can be identified as excitations of the Goldstone mode~\cite{suchanek2023irreversible}.}

{We can simplify even further by exploiting the fact that the dynamics of the model of \red{Eq.~\eqref{equ:1}} is well described by the lowest Fourier mode, alone~\cite{You_2020}.}
Applying a one mode approximation in Eq.~\eqref{eq:coupling} corresponds to neglecting all terms that are not multiples of $(\mathcal{A}_A^1)^3$. 
For the phase dynamics, this yields a decoupling of $\theta^1$ from higher modes, resulting in the approximate equation of motion
\begin{align}\label{equ:dyntheta1}
\dot {{\theta}}^1(t)=\mathbb{B} {{\theta}}^1+ \sqrt{\epsilon/L }\,{\zeta},
\end{align}
with ${{\theta}}^1 = ({\theta}_A^1, \tilde{\theta}_B^{1})^T$, $\tilde{\theta}_B^{1} \equiv\theta_B^1+\pi$,
\begin{align}\label{equ:operator}
\mathbb{B} = 
- \begin{pmatrix}
(\kappa^2-\delta^2)/\beta & -(\kappa^2-\delta^2)/\beta\\
-\beta & \beta\\
\end{pmatrix} \,,
\end{align}
and 
$\left\langle\zeta_i(t),\zeta_j(t)\right\rangle=(2/\mathcal{A}_{i}^{1,*})^2 \delta_{i,j}\delta(t-t')$, $i,j \in \{A,B\}$.
We present in App.~\ref{app:ap} the derivation of these closed-form expressions.
In App.~\ref{sec:numericalconf}, we provide a direct comparison between the analytical predictions by Eq.~\eqref{equ:dyntheta1} and numerical results, which shows that the 
dynamics of the first-mode phase ${{\theta}}^1$ is indeed accurately captured by the approximate expressions \red{Eqs.~\eqref{equ:dyntheta1} and \eqref{equ:operator}} (see Fig.~\ref{fig:delta_theta}). 
{This separation of the dynamics enables us to calculate the isolated contribution of $\theta^1$ to the entropy production, which we denote by $\sigma^1_{\theta}$. It can be interpreted as measure of TRSB due to relative and joint translations of the demixing profiles.
Since the emergence of the traveling wave phase can be characterized as an instability of $\theta_{A,B}^{1,*}$~\cite{You_2020}, this contribution is of particular interest.
}

Interestingly, the dynamics in \red{Eq.~\eqref{equ:dyntheta1}} is identical to the overdamped equations of motion of two individual degrees of freedom coupled by nonreciprocal, linear forces, each being coupled to its individual heat bath. Hence, we can
draw on the results of Ref.~\cite{loos2020irreversibility}, where a general expression for the entropy production of such a system is given. We find
\begin{align}\label{eq:theta1contribution}
     \sigma^1_\theta= 4q_1^2\,\chi_B\frac{\delta^2}{\delta_c^2-\delta^2}+\mathcal{O}(\epsilon)\,.
\end{align}
This suggests that the entropy production rate scales to leading order as $\mathcal{S}\sim\epsilon^{0}$, just as in the homogeneous phase; implying that $\mathcal{S}$ stays finite in the zero-noise limit. This is in line with our numerical results in Fig.~\ref{fig:scaling}. 
Furthermore, 
it implies that $\mathcal{S}$ does not scale merely quadratically with the nonreciprocity, different from the homogeneous phase [see Eq.~\eqref{res1}]. Specifically, $\sigma_\theta^1\sim \delta^2$ only holds for $\delta\to 0$, while it diverges like $\sigma_\theta^1\sim (\delta_c-\delta)^{-1}$ for $\delta\to \delta_c$; indicative of the cooperative character of the emerging dynamics. 

\begin{figure}
\includegraphics[width=.45\textwidth]{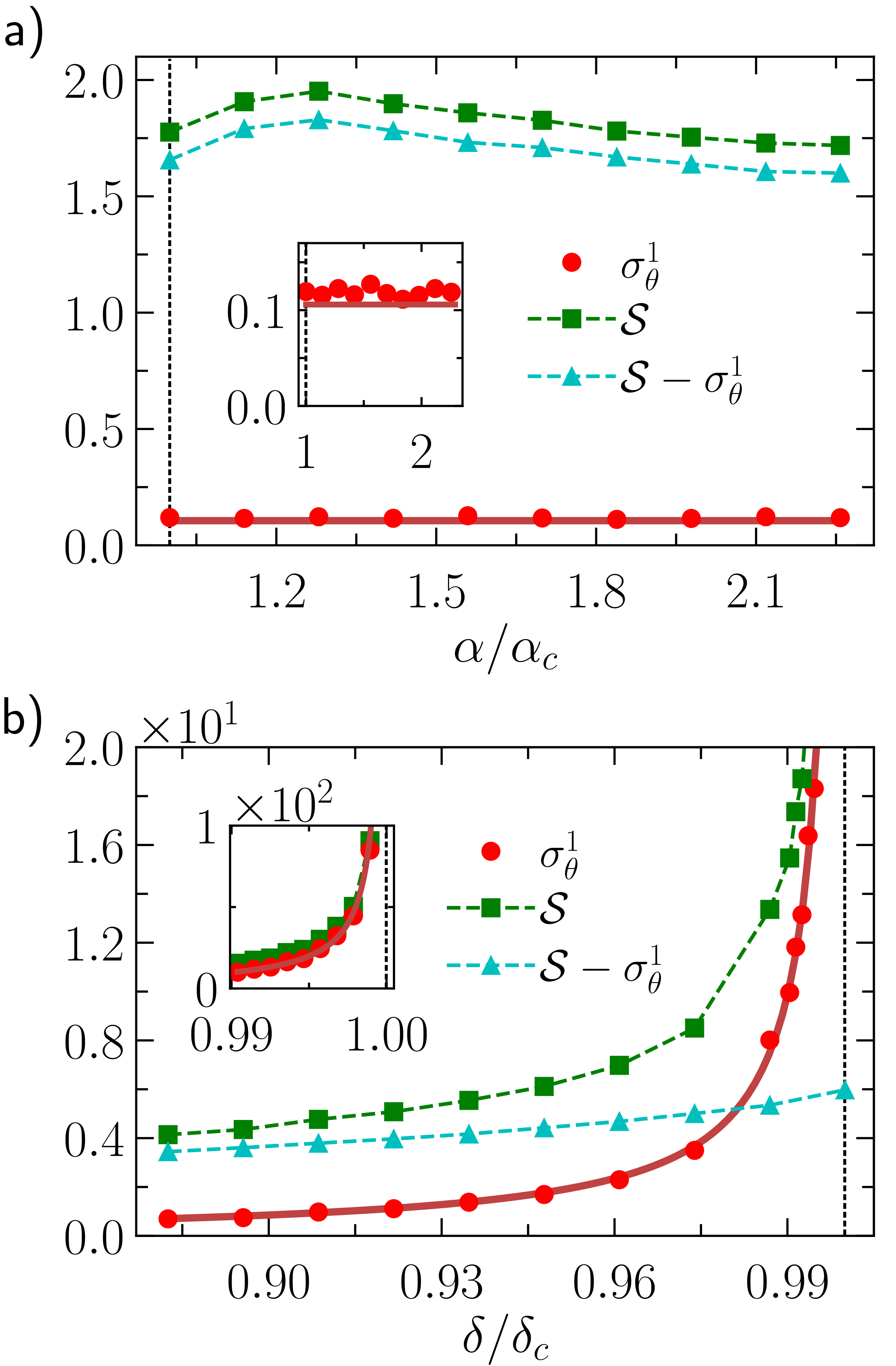}
\caption{\label{fig:stat}
\red{Total entropy production rate $\mathcal{S}$ from Eq.~(\ref{EPRnumtotal}), and its divergent contribution $\sigma^1_{\theta}$ due to the first-mode from Eqs.~(\ref{def:Spnum_k}),} as a function of the control parameters $\alpha$ and $\delta$, in the static-demixed phase: (a) near the transition to the mixed phase (here, $\delta=0.8\delta_c$, path $c_3$ from Fig.~\ref{fig:pd}), and (b) near the transition to the traveling-wave phase ($\alpha=-0.07$, path $c_4$ in Fig.~\ref{fig:pd}).
The analytical prediction \red{Eq.~\eqref{eq:theta1contribution}} is shown by solid red lines. Symbols depict simulation results with dashed lines as guides to the eye. Insets depict zooms into a smaller $y$ or $x$-axis ranges. 
We observe that $\mathcal{S}$ decreases
        away from the phase boundaries. For large $\vert\alpha\vert$, the data indicates saturation to a finite value. 
        Similarly to the homogeneous phase (Fig.~\ref{fig:hom}), $\mathcal{S}$ remains regular toward to static-static transition, while it surges toward the static-dynamic transition at $\delta=\delta_c$. 
        The decomposition of $\mathcal{S}$ clearly shows that the contribution $\sigma_\theta^1$ from the fluctuations of the first Fourier modes, ${{\theta}}^1 = ({\theta}_A^1, \tilde{\theta}_B^{1})^T$, is responsible for the surge in entropy production. It can exclusively be attributed to excitations of the Goldstone mode~\cite{suchanek2023irreversible}.
 Other parameters are $\kappa=0.01$, $\gamma=0.015$, $\beta=0.05$, $\epsilon=10^{-10}$.
}
\end{figure}

Thus, according to Eq.~\eqref{eq:theta1contribution}, $\sigma_\theta^1$
remains regular at the static-static transition to the homogeneous phase. In contrast, concerning the 
transition from the static-demixed to the traveling-wave phase for $\delta \to \delta_c$,
Eq.~\eqref{eq:theta1contribution}
predicts a surge of $\mathcal{S}$, driven by the phase dynamics of the first mode.  
These predictions are corroborated by  numerical results presented in Fig.~\ref{fig:stat}.  
As we show in Ref.~\cite{suchanek2023irreversible}, $\theta^1$ can in fact further be separated into two parts; where one of them (denoted $\theta_c$ in Ref.~\cite{suchanek2023irreversible}) can be identified as the dynamics along the Goldstone mode. Only this part contributes to the entropy production, such that $\sigma_\theta^1$ is a direct measure of TRSB in the Goldstone mode.

Further, since the fluctuations of all the amplitudes $\mathcal{A}^k_{A,B}$, as well as the phases of $k>1$, $\theta^{k>1}_{A,B}$,  with $k\in\{2n+1 : n\in \mathbb{N}_0\}$, remain bounded toward the transition~\cite{You_2020}, we expect also the contributions to $\mathcal{S}$ stemming from these sources to remain bounded. 
{Indeed, our numerical results confirm that $\mathcal{S}-\sigma_\theta^1$
remains regular and rather insensitive to the transition [see Fig.~\ref{fig:stat}(b)].}

\subsection{Dynamical traveling-wave phase}
\begin{figure}
\includegraphics[width=.45\textwidth]{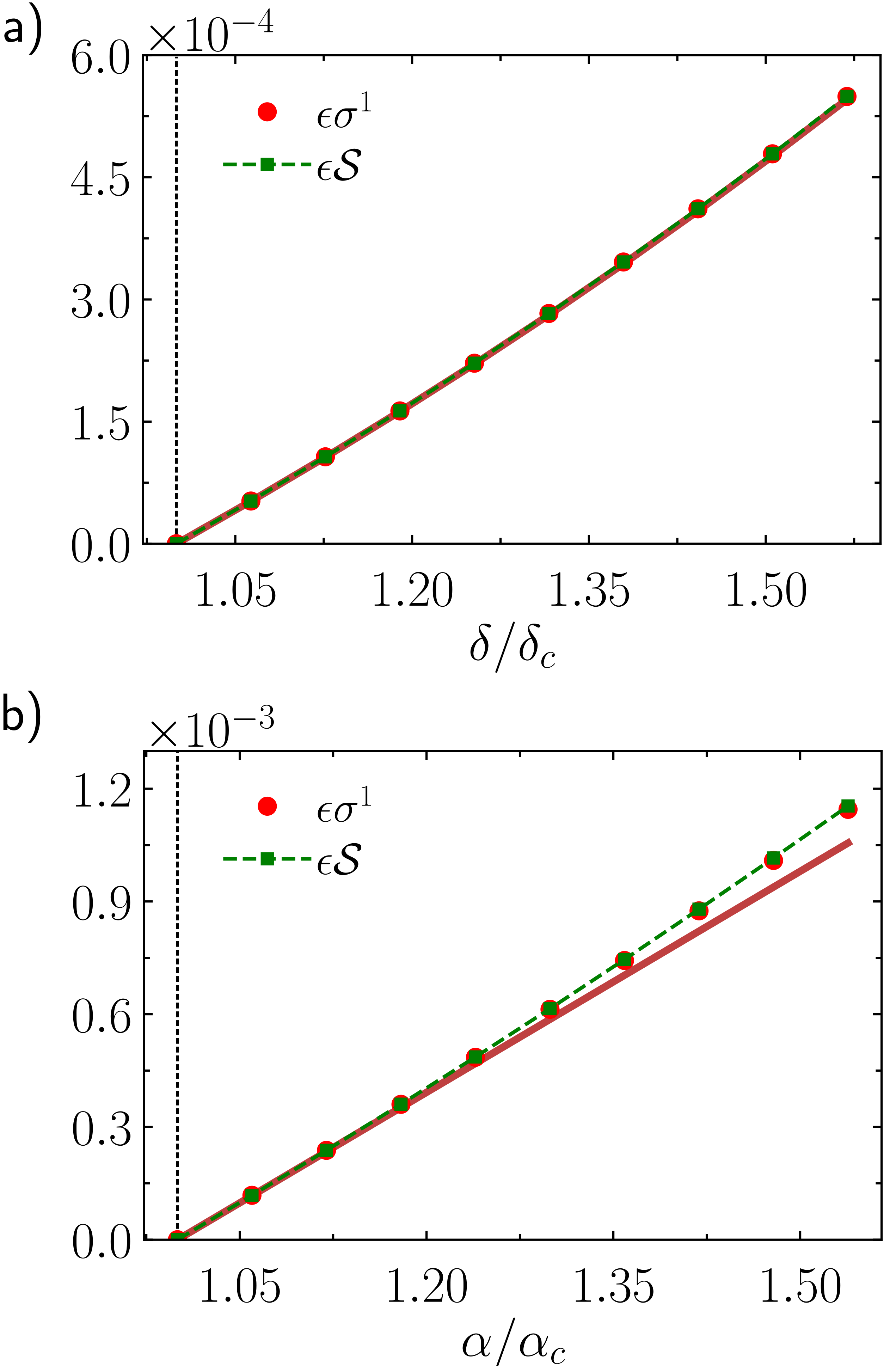}
\caption{
Total entropy production \red{$\mathcal{S}$ from Eq.~(\ref{EPRnumtotal}) and its first-mode's contribution $\sigma^1$ from Eq.~(\ref{EPRnumk})} within the traveling-wave phase in the vicinity of the phase boundaries: (a) toward the transition to the static-demixed phase [here, $\alpha=-0.07$, path $c_5$ in Fig.~\ref{fig:hom}(a)], and in (b) toward the mixed phase [here, $\delta=1.2\delta_c$, path $c_6$ in Fig.~\ref{fig:pd}(a)]. The analytical result for $\sigma^1$ \red{in Eq.~\eqref{eq:EPR-travelingwave}} is shown by solid red lines, symbols depict corresponding simulation results.
As one moves deeper into the traveling-wave phase, $\mathcal{S}$ generally increases, i.e., it increases with increasing amplitude of the traveling-wave $(\mathcal{A}^{1,*}_{A,B})^2\propto \vert\alpha-\alpha_c\vert$~\cite{You_2020}, as well as with an increase of the deterministic speed of the
traveling waves $\mathfrak{v}^2 \sim |\delta-\delta_c|$ [panel (a)].
}\label{fig:trav}
\end{figure}


Last, we turn to the traveling-wave phase itself. Here, already the deterministic part of the dynamics seems to exhibit a broken time-reversal symmetry. Indeed, the solutions of Eq.~\eqref{equ:1} in the limit $\epsilon\to 0$ assume the form ${\phi}^*_{A,B}(r,t)=\varphi_{A,B}(r\pm \mathfrak{v}t)$; which clearly break the $\mathcal{T}$-symmetry.
Thus, we expect $\mathcal{S}$ to contain contributions from the irreversibility of the fluctuations and from the noise-independent stationary motion of the profiles. 
In Ref.~\cite{suchanek2023connection} we show that the dominant contribution stemming from the deterministic motion is given by

\begin{align}\label{eq:EPR-travelingwave1}
    \mathcal{S}=\epsilon^{-1}\mathfrak{v}^2\int\limits_0^{L}\mathrm{d}r\big(\vert{\varphi_A}^*\vert^2+\vert{\varphi_B}^*\vert^2\big)+ \mathcal{O}(\epsilon^{0})\,.
\end{align}
From this expression, we can further derive an approximate solution for the leading-order contribution by the first mode.
Specifically, as shown in Ref.~\cite{You_2020}, the solutions in the traveling-wave phase approximately take the form ${\varphi}^*_{A,B}(t,r)=\mathcal{A}^{1,*}_{A,B}\cos (q_1 r+\mathfrak{v}t+\theta^{1,*}_{A,B} )$ with $\mathfrak{v} = \pm \sqrt{\delta^2-\delta_c^2}$ and $\mathcal{A}^{1,*}_{A}=2\sqrt{-\alpha-\gamma-\beta}$, 
$\mathcal{A}^{1,*}_{B}=\sqrt{(\delta+\kappa)/(\delta-\kappa)}\mathcal{A}^{1,*}_{A}$, which yields
\begin{align}\label{eq:EPR-travelingwave}
    \sigma^1=
   \epsilon^{-1}4L\frac{\delta(\delta^2-\delta_c^2)}{\kappa-\delta}(\alpha+\beta+\gamma)+ \mathcal{O}(\epsilon^{0})\,. 
\end{align}
According to Eq.~\eqref{eq:EPR-travelingwave},
$\mathcal{S}$ depends  \textit{linearly} both on $\alpha$ and $\delta$ above the transition, in excellent agreement with the numerical data shown in Fig.~\ref{fig:trav}. Equation~\eqref{eq:EPR-travelingwave} further implies that, in the low noise regime, 
$\mathcal{S}$ scales like $\sim\epsilon^{-1}$, i.e., it is dominated by the deterministic dynamics, which perfectly matches to the numerical data in Fig.~\ref{fig:scaling}. As a consequence, $\mathcal{S}^*$ diverges (independently of the choice of a UV cutoff). 
This ``physical divergence'' can easily be understood from Eq.~\eqref{eq:EPR-travelingwave1}, which makes apparent that, in the low noise regime, $\mathcal{S}$ simply mirrors the presence of the deterministic ``mass" current $\mathfrak{v}{\varphi}^*_{A,B}$.


\section{Discussion}

We have studied a model of two noisy Cahn-Hilliard fields with linear, asymmetric cross-coupling~\cite{You_2020,Saha_2020,frohoff2021suppression}, as a prototypical example of a nonreciprocal active field theory. It admits the emergence of a dissipative order parameter pattern in the form of a traveling wave. The dynamical phase can be entered via two different static-dynamic phase transitions, an oscillatory instability or a critical exceptional point, respectively, which are both characteristic for many-body systems with nonreciprocal interactions~\cite{Fruchart2021}. 

For the three distinct phases of the model (homogeneous, static-demixed, and traveling-wave phase), we found 
 perturbative analytical expressions for the irreversibility measure $\mathcal{S}$. For the homogeneous phase and for the traveling-wave phase, these become asymptotically exact in the limit of vanishing noise intensity. For the static-demixed phase, we had to refine the approach and make some additional simplification to tackle the problem analytically. The approximate results thereby obtained are indeed in excellent agreement with our numerical data, especially near the phase boundaries.

In the vicinity of the phase transitions, $\mathcal{S}$ essentially exhibits two different types of behavior.
Along the transition line connecting the two static phases of the model, we find that $\mathcal{S}^*$ is finite and continuous, despite a diverging susceptibility, at the transition~\cite{suchanek2023connection}. In contrast, across both static-dynamic transitions, we observe a massive increase in entropy production that already starts in the static phases. This is particularly interesting in view of the fact that the respective static phases themselves (homogeneous and static-demixed) appear, at first glance, indistinguishable from the corresponding equilibrium phases. The surge of entropy production means that the formation of the dynamical phase is heralded by a gradual, yet drastic increase of irreversible mesoscale dynamics; a feature which would be hidden when looking at the corresponding zero-temperature model. 
These numerical and analytical results for the nonreciprocal Cahn-Hilliard model corroborate and illustrate our general predictions for the TRSB across the continuous phase transitions of a broad class of non-Hermitian field models, presented in the two companion papers~\cite{suchanek2023irreversible,suchanek2023connection}. 

Additionally, our explicit perturbative calculations for the nonreciprocal Cahn-Hilliard model shed light on the origin and scale-dependence of TRSB throughout the phase diagram. Specifically, we could attribute the surge of entropy production in both static phases exclusively to the 
contribution of the first Fourier mode.
This means that, in this model, most of the TRSB occurs through the 
coherently activated long-wavelength dynamics,
as might have been expected for an emerging dissipative mesoscale pattern.

Our analytical considerations moreover revealed the concrete dependence of the entropy production
on the strength of nonreciprocity, $\delta$, throughout the phase diagram. This aspect deserves some further discussion.
First, we found that in the homogeneous phase, $\mathcal{S}\sim \delta^2$ [see Eq.~\eqref{res1}]. In contrast, in the static-demixed phase, $\mathcal{S}\sim \delta^2$ holds only close to equilibrium ($\delta=0$). Near the transition to the dynamical phase, the contribution to leading order in $\epsilon$ scales as $(\delta_c-\delta)^{-1}$ [see Eq.~\eqref{eq:theta1contribution}], revealing a threshold mechanism for wave propagation.
Finally, in the dynamical phase itself, close to the transition, $\mathcal{S}$ depends linearly on $\delta-\delta_c$, to leading order in $\epsilon$ [Eq.~\eqref{eq:EPR-travelingwave}]. Further away from the transition, $\mathcal{S}\sim \delta^2$.
Thus, in the low noise \red{regime}, the $\delta$-dependence of $\mathcal{S}$ is phase sensitive. 
Notably, this interesting $\delta$-dependence only becomes apparent when a suitable regularization (UV cutoff) is introduced. Otherwise, the relevant properties of $\mathcal{S}$ are obscured by trivial but overwhelming contributions of spurious (and entirely redundant) degrees of freedom [see Eq.~\eqref{eq:sigmainf} and App.~\ref{sec:eprhom}].

A particularly interesting insight gained from the analytical treatment concerns the TRSB in the static-demixed phase.
{Specifically, we could show that the main part of the entropy production stems for the fluctuations that lead to displacements of the demixing profiles (rather than their distortion).}
Building upon our expressions \red{Eq.~\eqref{equ:dyntheta1}} and \red{Eq.~\eqref{equ:operator}}, we could further formally map the interface dynamics  to the {dissipative ballistic} motion of an \red{\textit{active Ornstein-Uhlenbeck particle}}, providing a vivid {and palpable mechanistic particle interpretation} 
{for the TRSB in the active field theory~\cite{suchanek2023irreversible}.} 
{Since the active interface dynamics resides in a Goldstone mode of the model, one may say that it sacrifices its mesoscale {time-translation} invariance in order to restore the broken spatial (translational) symmetry more effectively.}
\red{This ``accumulation'' of TRSB in the first mode, and its active-particle interpretation, could also be subsumed under the term 
``active Goldstone modes".}

Finally, it is interesting to compare  the scaling of $\mathcal{S}$ in the noise intensity to other active field theories. 
Earlier works~\cite{Nardini2017,Li_2021,Borthne2020} have studied it in systems that display activity-driven pattern formation, such as motility-induced phase separation in active model $B$ or $B^+$~\cite{Nardini2017,Li_2021}, and a flocking phase in polar active models~\cite{Borthne2020}. In all of these models, it was found that $\mathcal{S}$ scales like $\sim \epsilon^{1}$ within the homogeneous phases, implying time-symmetric, equilibrium-like dynamics in the limit $\epsilon \to 0$. Only in the phase-separated~\cite{Nardini2017,Li_2021} or polar~\cite{Borthne2020} state, $\mathcal{S}$ scales like $\sim \epsilon^{0}$ and hence remains finite in the zero-noise limit.
In contrast, we found that, in the presence of nonreciprocal coupling, $\mathcal{S}$ remains nonzero in the noise-free limit throughout the whole phase diagram, including the homogeneous phase, revealing that the nonreciprocal coupling is a qualitatively different (inherent) source of irreversibility.

\red{In future studies, it could be interesting to extend the investigation of noise-induced interface motion, observed in the one-dimensional nonreciprocal Cahn-Hilliard model, to higher dimensions where more complex interface geometries could arise.} 
Another perspective would be the extension of our considerations to more complex models, such as mesoscopic descriptions for active-passive mixtures~\cite{Wittkowski_2017}, where a nonreciprocal coupling between the active and passive coarse-grained fields naturally emerges. {More generally, one could explore the range of possibilities for concrete atomistic realizations that can, on the mesoscale, be subsumed into the studied nonreciprocal Cahn-Hilliard phenomenology. In particular, it could be interesting to see whether these comprise both cases in which the dynamic symmetry breaking arises on the single particle level \cite{Wang23} and on the many-body level \cite{Fruchart2021}, respectively.} Another possible direction could be to study thermodynamic embeddings, similar to the one described in Ref.~\cite{Cates2021}, and to complement the model by a field that fuels the TRSB fluctuations.

{We note that simultaneously and independently of ours, a recent study~\cite{alston2023irreversibility} reported consistent results on the TRSB in the nonreciprocal Cahn-Hilliard model.}

\vspace*{1cm}
\begin{acknowledgments}
SL acknowledges funding by the Deutsche Forschungsgemeinschaft (DFG, German Research Foundation) – through the project 498288081.
TS acknowledges financial
support by the pre-doc award program at Leipzig University. 
SL thanks the Physics Institutes of Leipzig University for their hospitality during several research stays.

\end{acknowledgments}

\appendix





\section{Coupling of different modes in the demixed state}\label{app:ap}

In this appendix, we outline our approach to tackle analytically the fluctuations in the static-demixed phase, described in Sec.~\ref{sec:static-demixed}.
Specifically, we here show that the stochastic dynamics can be separated into two parts: fluctuations of the amplitude $\{\mathcal{A}^k\}$ with odd $k\in\{2n+1 : n\in \mathbb{N}_0\}$, as well as all fluctuations of $\phi^k_{A,B}$ for even $k$, and second, fluctuations of $\{\theta^j\}$ which encode the dispersion and displacements of the demixing profiles, see Sec.~\ref{sec:static-demixed}. Furthermore, concerning the second part, we show that 
within the one mode approximation, the contributing stochastic dynamics can be represented by closed-form equations for $\theta^1$. In the following, we will show step by step how to achieve these closed-form expressions.

First, we consider $k\in\{2n+1 : n\in \mathbb{N}_0\}$ and derive the equations of motion for $\mathcal{A}^k$ and $\mathcal{\theta}^k$. To this end, we Fourier transform Eq.~\eqref{equ:1} and then apply the It\^{o} formula to
	\begin{align}
		\theta_{A,B}^k&=\arctan\frac{\mathrm{Im}{\phi}_{A,B}^k}{\mathrm{Re}{\phi}_{A,B}^k}\,,\\\nonumber
		\\
		\mathcal{A}_{A,B}^k&=2\sqrt{\mathrm{Re}{\phi}_{A,B}^k+\mathrm{Im}{\phi}_{A,B}^k}\,.
	\end{align}
	Noting that the quadratic variations for the amplitude and the phase processes are determined by
	\begin{align}
		\nabla^2 \sqrt{x^2+y^2}=\frac{1}{\sqrt{x^2+y^2}}\,,\\
		\nabla^2 \arctan\left(\frac{x}{y}\right)=0\,,
	\end{align}
	we thereby obtain the respective stochastic equations of motions for the amplitude and the phase dynamics, which read
  \begin{widetext}
	\begin{align}\label{equ:dynex}
		\dot{\mathcal{A}}_A^k=&-q_k^2\left[(\alpha + \gamma q_k^2) \mathcal{A}_A^k+(\kappa-\delta)\mathcal{A}_B^k\cos(\theta_A^k-\theta_B^k)+\mathrm{Re}(\mathrm{K}^ke^{-i\theta_A^k})-\frac{4\epsilon}{L\mathcal{A}_A^k}\right]+\xi_{\mathcal{A}_A^k}\,,\\
		\dot{\mathcal{A}}_B^k=&-q_k^2\left[\beta \mathcal{A}_B^k+(\kappa+\delta)\mathcal{A}_A^k\cos(\theta_A^k-\theta_B^k)-\frac{4\epsilon}{L\mathcal{A}_B^k}\right]
		+\xi_{\mathcal{A}_B^k}\,,\\
		\dot{\theta}_A^k=&\,q_k^2\left[\sin(\theta_A^k-\theta_B^k)\frac{\mathcal{A}_B^k}{\mathcal{A}_A^k}(\kappa-\delta)-
		q_k^2\mathrm{Im}(\mathrm{K}^ke^{-i\theta_A^k})\frac{1}{\mathcal{A}_A^k}\right]
		+\xi_{\theta_A^k}\,,\\
		\dot{\theta}_B^k=&-q_k^2\sin(\theta_A^k-\theta_B^k)\frac{\mathcal{A}_A^k}{\mathcal{A}_B^k}(\kappa+\delta)+\xi_{\theta_B^k}\,,
	\end{align}
	with
 \begin{align}\label{coupling term}
    \mathrm{K}^k= \sum\limits_{l,l'}\frac{q_{l}^2+2q_{l}q_{l'}}{4q_k^2}
    {\mathcal{A}^k_A\mathcal{A}^l_A}\mathcal{A}_A^{k-l'-l}
    e^{i(\theta_A^l+\theta_A^{l'}+\theta_A^{k-l'-l}-{i\theta_A^k})}.
\end{align}
 \end{widetext}

	For the transformed noise terms, we find
	\begin{align}
		\xi_{\mathcal{A}_{A,B}^k}=4\sqrt{\frac{\epsilon}{L}}\frac{\mathrm{Re}{\phi}^k_{A,B}\mathrm{Re}\xi_{A,B}^k+\mathrm{Im}{\phi}^k_{A,B}\mathrm{Im}\xi_{A,B}^k}{\mathcal{A}_{A,B}^k}\,,\\\nonumber
		\\
		\xi_{\theta_{A,B}^k}=4\sqrt{\frac{\epsilon}{L}}\frac{\mathrm{Re}{\phi}^k_{A,B}\mathrm{Im}\xi_{A,B}^k-\mathrm{Im}{\phi}^k_{A,B}\mathrm{Re}\xi_{A,B}^k}{(\mathcal{A}_{A,B}^k)^2}\,,
	\end{align}
	such that the co-variance matrix again turns out to be diagonal. The transformation has rendered the noise  multiplicative. The diagonal elements of its co-variances  matrix are given by
 \begin{align}
     \big\langle\xi_{\mathcal{A}_{A}^k}(t)\xi_{\mathcal{A}_{A}^l}(t')\big\rangle&=\big\langle\xi_{\mathcal{A}_{B}^k}(t)\xi_{\mathcal{A}_{B}^l}(t')\big\rangle=q_k^2\frac{4\epsilon}{L}\delta_{kl} \, ,
     \\
     \big\langle\xi_{\theta_{A}^k}(t)\xi_{\theta_{A}^l}(t')\big\rangle&=q_k^2\frac{4\epsilon}{L\mathcal{A}_A^2}\delta_{kl} \, ,
     \\
     \big\langle\xi_{\theta_{B}^k}(t)\xi_{\theta_{B}^l}(t')\big\rangle&=q_k^2\frac{4\epsilon}{L\mathcal{A}_B^2}\delta_{kl}\, .
 \end{align}
We parametrize the solutions of Eq.~\eqref{equ:1} for $\epsilon=0$ as
\begin{align}
\phi^*_{A,B}(r,t)=\sum\limits_{k>0}\mathcal{A}^{k,*}_{A,B}\cos\left[q_k r+\theta^{k,*}_{A,B}(t)\right],
\end{align}
with wavenumbers $q_k\equiv 2\pi k /L$, and odd $k \in \{2n+1:n\in\mathbb{N}_0\}$.
 Since the solutions in the static-demixed state are parity symmetric \cite{You_2020}, we can always center our $r$-axis such that, $\theta_{A,B}^{k,*}\in \{0,\pi\}$.
We define
	\begin{align}\nonumber
		\Delta\theta_{A,B}^k&=\theta_{A,B}^{k}-\theta_{A,B}^{k,*}\,,\\
		\Delta\mathcal{A}_{A,B}^k&=\mathcal{A}_{A,B}^k-\mathcal{A}_{A,B}^{k,*}\,,
	\end{align}
	and therewith find that, up to linear order in $\{\Delta\theta_{A,B}^k\}$ and $\{\Delta\mathcal{A}_{A,B}^k\}$, 
 \out{
 ,
     
	\begin{align}\nonumber
		\mathrm{Im}\big(\mathrm{K}^ke^{-i\theta_A^k}\big)=&\frac{1}{4}
		\sum\limits_{l,l'}\big(q_{l}^2+2q_{l}q_{l'}\big)\mathcal{A}^{l}_A\mathcal{A}^{l'}_A\mathcal{A}_A^{k-{l}-{l'}}
  \times
  \nonumber \\
  &
  \sin\big(\theta_A^{l}+\theta_A^{l'}+\theta_A^{k-{l}-{l'}}-\theta_A^k\big)\,,
	\end{align}
 }
 \begin{widetext}
	\begin{align} \label{equ:imK}
		\frac{\mathrm{Im}\big(\mathrm{K}^ke^{-i\theta_A^k}\big)}{\mathcal{A}^{k}}=&~\frac{1}{4}
		\sum\limits_{{l},{l'}}(q_{l}^2+2q_{l}q_{l'})\frac{\mathcal{A}^{l,*}_A\mathcal{A}^{l',*}_A\mathcal{A}_A^{k-{l}-{l'},*}}{\mathcal{A}_A^{k,*}}
		\big(\Delta\theta_A^{l}+\Delta\theta_A^{l'}+\Delta\theta_A^{k-{l}-{l'}}-\Delta\theta_A^k\big)g^{k,l,l'}
		\, ,
  \\
		\mathrm{Re}(\mathrm{K}^k e^{-i\theta_A^k})=&~\frac{1}{4}\sum\limits_{{l},{l'}}(q_{l}^2+2q_{l}q_{l'})\left(\mathcal{A}^{l,*}_A\mathcal{A}^{l',*}_A \Delta\mathcal{A}_A^{k-{l}-{l'}}+\mathcal{A}^{l,*}_A\mathcal{A}_A^{1-{l}-{l'},*}\Delta\mathcal{A}^{l'}_A+\mathcal{A}^{l',*}_A\mathcal{A}_A^{1-{l}-{l'},*}\Delta\mathcal{A}^{l}_A\right)h^{k,l,l'}
		\, , \label{equ:reK}
	\end{align}
  with $g^{k,l,l'}=\cos\big(\theta_A^{l,*}+\theta_A^{l',*}+\theta_A^{k-{l}-{l'},*}-\theta_A^{k,*}\big)$
and 
  \begin{align}
		\sin\big(\theta_A^k-\theta_B^k\big)\frac{\mathcal{A}_B^k}{\mathcal{A}_A^k}=&\cos\big(\theta_{A}^{k,*}-\theta_{B}^{k,*}\big)\big(\Delta\theta_A^{k}-\Delta\theta_B^k\big)\frac{\mathcal{A}_B^{k,*}}{\mathcal{A}_A^{k,*}}
		\, , \\
		\sin\big(\theta_A^k-\theta_B^k\big)\frac{\mathcal{A}_A^k}{\mathcal{A}_B^k}=&\cos\big(\theta_{A}^{k,*}-\theta_{B}^{k,*}\big)\big(\Delta\theta_A^k-\Delta\theta_B^k\big)\frac{\mathcal{A}_A^{k,*}}{\mathcal{A}_B^{k,*}}\, .
  \end{align}
   \end{widetext}
Inserting this into Eq.~\eqref{equ:dynex}, we see that in the \red{regime} of low noise intensity, the dynamics of the amplitude and the phase are decoupled.
Further, from $\mathcal{A}_{A,B}^{k,*}=0$ for $k\in\{2n : n\in \mathbb{N}_0\}$, we can conclude that all upper indices in Eqs.~\eqref{equ:imK} and \eqref{equ:reK} are uneven. 
Hence, the dynamics of $\{\Delta\theta_{A,B}^k\}$ and $\{\Delta\mathcal{A}_{A,B}^k\}$ for $k\in\{2n+1 : n\in \mathbb{N}_0\}$ is decoupled from the dynamics of ${\phi}^k$ for $k\in\{2n : n\in \mathbb{N}_0\}$. Vice versa, expanding the coupling term \red{Eq.~\eqref{coupling term}} for even $k$, results only in coupling among even wavenumbers.  Therefore, combining all of these results, we find that the dynamics ${\phi}^k$ for $k\in\{2n+1 : n\in \mathbb{N}_0\}$ and ${\phi}^k$ for $k\in\{2n : n\in \mathbb{N}_0\}$ are mutually decoupled in the low noise regime.  

Next, for $|k|>1$ we redefine
	\begin{align}
		\Delta\theta_A^k\rightarrow\Delta\theta_A^k-\frac{q_k}{q_1}\Delta\theta^1_A \, .
	\end{align}
	Then, using ${\mathcal{A}_B^{k,*}}/{\mathcal{A}_A^{k,*}}= ({\kappa+\delta})/{\beta}$ \cite{You_2020} and $\theta_{A}^{1,*}=0$,  $\theta_{B}^{1,*}=\pi$, the equations of motion for the first mode read 
	\begin{equation}\label{equationinter}
	 \begin{aligned} 
		\partial_t{\Delta\theta}_A^1=&-q_k^2\big(\Delta\theta_A^1-\Delta\theta_B^1\big)\frac{\kappa^2-\delta^2}{\beta}+ Q+\xi_{\theta_A^1}\,,\\ 
	\partial_t{\Delta\theta}_B^1=&~q_k^2\big(\Delta\theta_A^1-\Delta\theta_B^1\big)\beta+\xi_{\theta_B^1}\,,
	\end{aligned}
    \end{equation}
	where the term  $Q\big(\{\Delta\theta_A^k\}_{k>1}\big)$, coupling different Fourier modes, is given by
 \begin{widetext}
	\begin{align}\label{theta_app}
		Q\big(\{\Delta\theta_A^k\}_{k>1}\big)
		=&~
		\frac{1}{4}\sum\limits_{k,l}\big(q_k^2+2q_kq_l\big)\frac{\mathcal{A}^{k,*}_A\mathcal{A}^{l,*}_A\mathcal{A}_A^{1-k-l,*}}{\mathcal{A}_A^{1,*}}\big(\Delta\theta_A^k+\Delta\theta_A^l+\Delta\theta_A^{1-k-l}\big)\,.
	\end{align}
 \end{widetext}
	Now we use that in a wide range of the phase diagram it holds that $\mathcal{A}^{k,*}_A\ll \mathcal{A}^{1,*}_A$ \cite{You_2020}. After careful consideration of all coefficients in Eq.~\eqref{theta_app}, we find that the most relevant one is given by $2\mathcal{A}^{1,*}_A\mathcal{A}^{3,*}_A$. Since the other coefficients in Eq.~\eqref{equationinter} are $\mathcal{O}\big((\mathcal{A}^{1,*}_A)^2\big)$, we conclude that the term $Q$ as a whole will have minor influence of on the dynamics of $\Delta\theta^1_{A,B}$ and can therefore approximately be neglected. This leads to a closed equation of the form of Eq.~\eqref{equ:dyntheta1}. Indeed, our numerical results in App.~\ref{sec:numericalconf} show
	that this approximation accurately capture the properties of the exact solution.

\section{Numerical evaluation of the entropy production rate}\label{sec:num}

Here we show how the steady-state entropy production rate $\mathcal{S}$ can be inferred from the numerical simulation of Eq.~\eqref{equ:1}. 
For this, we used the classical Euler-Mayurama algorithm \cite{platen2010numerical} with finite difference gradients on a domain of length $L=2\pi$, discretized by $32$ equally spaced mesh points and time slices of $\Delta t= 0.01$. 

For $\phi^*=0$ the numerical entropy production rate is simply obtained by inserting the numerical results for $\mathrm{Cov}(\phi)$ into Eq.~\eqref{eq:EPRmodes}.
For $\vert\phi^*\vert >0$ however, we have to apply a different approach since we want to access the contributions assigned to the amplitude and the phase dynamics, separately. 
 This can be achieved as follows.
Along the same lines as in App.~\ref{app:ap}, we find that the equation of motion for $\theta^k$ read
	\begin{align}
		\dot{\theta}_A^k=&k^2\sin(\theta_A^k-\theta_B^k)\frac{\mathcal{A}_B^k}{\mathcal{A}_A^k}(\kappa-\delta)+H^k+\xi_{\theta_A^k}\,,\\
		\dot{\theta}_B^k=&-k^2\sin(\theta_A^k-\theta_B^k)\frac{\mathcal{A}_A^k}{\mathcal{A}_B^k}(\kappa+\delta)+\xi_{\theta_B^1}\,,
	\end{align}
	with
	\begin{align}
		H^k=&\frac{1}{4}\sum\limits_{{l},{l'}}({l}^2+2{l}{l'})\mathcal{A}^{l}_A\mathcal{A}^{l'}_A\mathcal{A}_A^{k-{l}-{l'}}
  \nonumber \\
  &\times
  \sin\big(\theta_A^{l}+\theta_A^{l'}+\theta_A^{k-{l}-{l'}}-\theta_A^k\big)\,.
	\end{align}
	Analogously,
	we find the equations of motion for the amplitude part 
	\begin{align}
		\dot{\mathcal{A}}_A^k=&-k^2\bigg[\!(\alpha + \gamma k^2) \mathcal{A}_A^k+(\kappa-\delta)\mathcal{A}_B^k-\frac{\epsilon}{\mathcal{A}_A^k}+J^k\bigg]
  \nonumber \\
  &+\xi_{\mathcal{A}_A^k}\,,\\
		\dot{\mathcal{A}}_B^k=&-\bigg[\beta \mathcal{A}_B^k+(\kappa+\delta)\mathcal{A}_A^k-\frac{\epsilon}{\mathcal{A}_B^k}\bigg]+\xi_{\mathcal{A}_B^k}\,,
	\end{align}
	with
	\begin{align}
		J^k=&\frac{1}{4}\sum\limits_{{l},{l'}}({l}^2+2{l}{l'})\mathcal{A}^{l}_A\mathcal{A}^{l'}_A\mathcal{A}_A^{k-{l}-{l'}}
  \nonumber \\
  &
  \times\cos\left(\theta_A^{l}+\theta_A^{l'}+\theta_A^{k-{l}-{l'}}-\theta_A^k\right)\,.
	\end{align}    
	Applying the standard procedure for the computation of the entropy production rate of Langevin processes outlined in detail in \cite{Seifert2005, Li_2021},
	we find that the entropy production assigned to a section of the trajectory $\{\mathcal{A}(t),\theta(t)\}_{t\in [0,+\infty)}$ is given by
 \begin{widetext}
	\begin{align}
		 s[\mathcal{A},\theta,t,t+T]=~&
   \frac{\pi \delta}{\epsilon}
    \sum\limits_k\bigg[
		\int\limits_t^{t+T}\mathrm{d}t\,\mathcal{A}^k_A\mathcal{A}_B^k\big(\dot{\theta}_A^k+\dot{\theta}_A^k\big)\sin\big(\theta^k_A-\theta^k_B\big)+\int\limits_t^{t+T}\mathrm{d}t\,\big(\dot{\mathcal{A}}_B^k\mathcal{A}^k_A-\dot{\mathcal{A}}_A^k\mathcal{A}_B^k\big)\cos\big(\theta^k_A-\theta^k_B\big)\bigg]
 \, . 
	\end{align}
  \end{widetext}
 Therefore, the contribution to  $\mathcal{S}$ associated with the dynamics of the phase can be computed  by numerically evaluating \red{the  averages
 \begin{widetext}
 \begin{align}\label{def:Spnum_k}
     \sigma_\theta^\mathrm{k,num}={\pi}\epsilon^{-1}\delta
		\mathcal{A}^k_A\mathcal{A}_B^k\left[\left\langle\frac{\theta^k_A(t+\Delta t)-\theta^k_A(t)}{\Delta t}\sin\big(\theta^k_A-\theta^k_B\big)\right\rangle+\left\langle\frac{\theta^k_B(t+\Delta t)-\theta^k_B(t)}{\Delta t}\sin\big(\theta^k_A-\theta^k_B\big)\right\rangle\right]
 \end{align}
 \end{widetext}
 for each mode, resulting in the total phase contribution
 \begin{align}\label{def:Spnum}
		\mathcal{S}_\theta^\mathrm{num}=\sum\limits_k
  \sigma_\theta^\mathrm{k,num}\,.
\end{align}
Analogously,  we find that the contribution to $\mathcal{S}$ associated with the dynamics of the amplitude is given by
 \begin{align}\label{def:SAnum}
		\mathcal{S}_\mathcal{A}^\mathrm{num}=\sum\limits_k
 \sigma_\mathcal{A}^\mathrm{k,num}\,,
\end{align}
with
 \begin{widetext}
\begin{align}\label{def:SAnum_k}
		\sigma_\mathcal{A}^\mathrm{k,num}={\pi}\epsilon^{-1}\delta\left[\left\langle\frac{ \mathcal{A}^k_B(t+ t)- \mathcal{A}^k_B(t)}{\Delta t} \mathcal{A}^k_A\cos(\theta_A^k-\theta_B^k)\right\rangle-\left\langle\frac{ \mathcal{A}^k_B(t+ t)- \mathcal{A}^k_B(t)}{\Delta t} \mathcal{A}^k_A\cos(\theta_A^k-\theta_B^k)\right\rangle\right]\,.
	\end{align}
 \end{widetext}
 }
Consequently 
	\begin{align}\label{EPRnumk}
\sigma^\mathrm{k,num}=\sigma^\mathrm{k,num}_\mathcal{A}+\sigma^\mathrm{k,num}_\theta 
	\end{align}
 and
  \begin{align}\label{EPRnumtotal}
     \mathcal{S}^{\mathrm{num}}=\sum\limits_{k} \sigma^\mathrm{k,num}
 \end{align}
 The averages appearing in Eqs.~\eqref{def:Spnum_k} and \eqref{def:SAnum_k} are directly evaluated from the simulations using at least $10^6$ noise realizations. For the simulations shown in Figs.~\ref{fig:hom}, \ref{fig:scaling}, \ref{fig:stat} and \ref{fig:trav}, we have chosen the numerical cutoff wavenumber $k=5$ to characterize the behavior of $\mathcal{S}$ at the transitions. We tested higher cutoffs $k$ and obtained consistent results.

\section{Numerical verification of the one mode approximation}\label{sec:numericalconf}
\begin{figure}
\includegraphics[width=.45\textwidth]{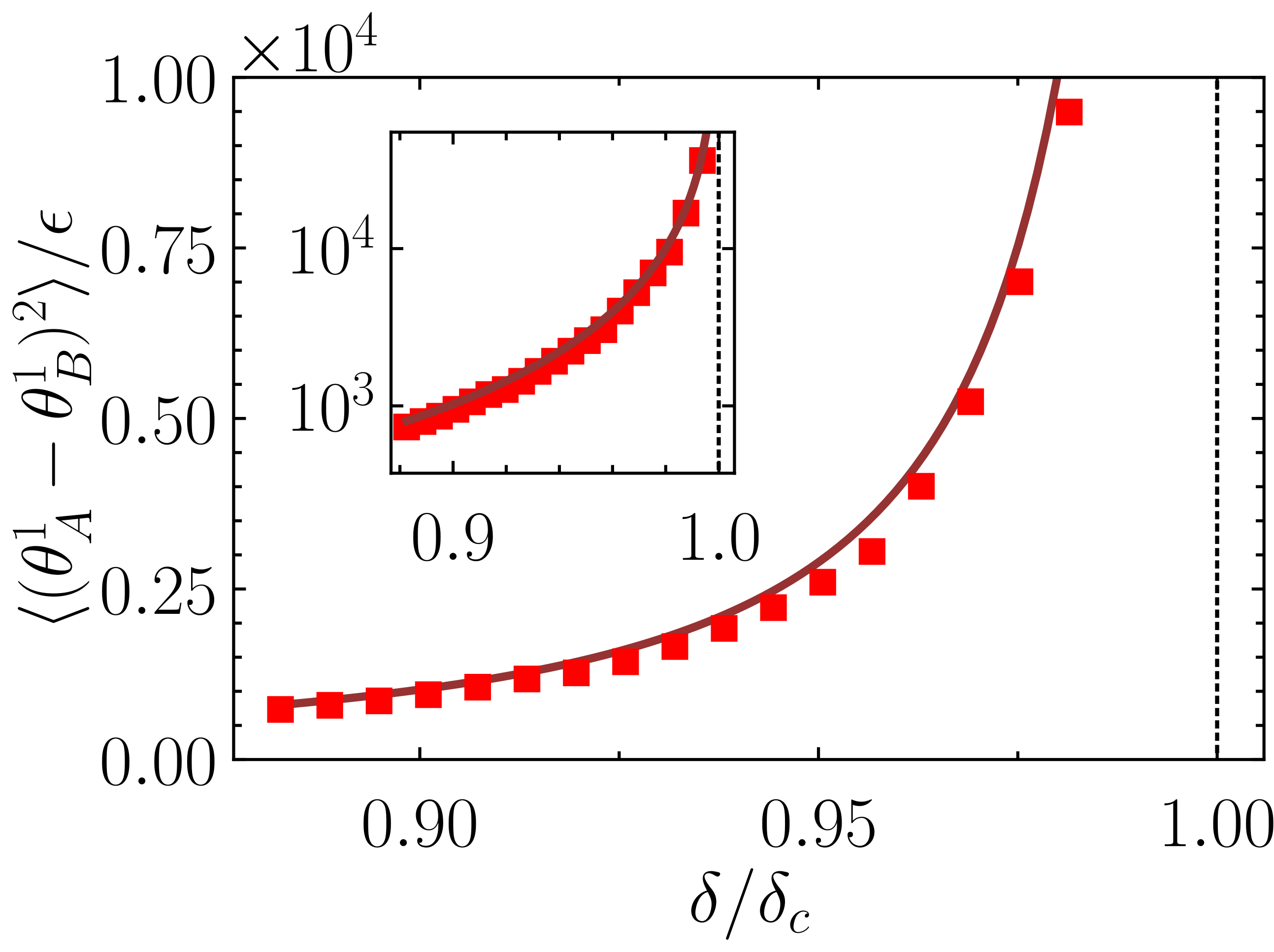}
\caption{\label{fig:delta_theta} 
Scaled variance of the phase fluctuations near the transition
as a function of $\delta$ for $\alpha=-0.07$, $\beta=0.05$, $\gamma=0.015$, $\kappa=0.01$, $\epsilon=10^{-10}$ from the numerical simulation of Eq.~\eqref{equ:1}. The solid line shows the analytical result Eq.~\eqref{theta_analytical} which holds within the one mode approximation. The inset shows the same results in a log-scale.}
\end{figure}
To evaluate the validity of the closed  form approximate dynamical equation for  ${\theta}^1$, Eq.~\eqref{equ:dyntheta1}, we consider the variance of the phase fluctuations $\big\langle ({\theta}_A^1- \tilde{\theta}_B^{1})^2\big\rangle$.
We evaluate this average numerically on a domain of length $L=2\pi$ using Eq.~\eqref{equ:dyntheta1} and compare it to the analytical prediction, which is obtained as follows.
Defining $\Delta\theta^\pi\equiv{\theta}_A^1- \tilde{\theta}_B^{1}$ and rearranging Eq.~\eqref{equ:dyntheta1}, we find
\begin{align}\label{deltaOU}
    \partial_t{\Delta\theta^\pi} &=
    -\frac{(\delta_c^2-\delta^2)}{\beta}\Delta\theta^\pi +\sqrt{\frac{2\epsilon({\Gamma}_A+{\Gamma}_B)}{\Gamma_A\Gamma_B}}\xi_{\Delta}\,,
\end{align}
with 
\begin{align}\Gamma_{A,B}\equiv\int_0^{2\pi} \mathrm{d}{r}\,\vert\phi^{1,*}_{A,B}({r})\vert^2=\pi\vert \mathcal{A}_{A,B}^{1,*}\vert^2\, ,
\end{align}
and
$\langle \xi_\eta(t)\xi_\nu(t')\rangle=\delta_{\eta\nu}\delta(t-t')$.
Noting that Eq.~\eqref{deltaOU} is an Ornstein-Uhlenbeck process \cite{Uhlenbeck30}, we find
\begin{align}\label{theta_analytical}
   \left\langle ({\theta}_A^1- \tilde{\theta}_B^{1})^2\right\rangle= \epsilon{\frac{{\Gamma}_A+{\Gamma}_B}{\Gamma_A\Gamma_B}}\frac{\beta}{\delta_c^2-\delta^2}\, ,
\end{align}
with
\begin{align}
    {\frac{{\Gamma}_A+{\Gamma}_B}{\Gamma_A\Gamma_B}}=\frac{\beta^2+(\delta+\kappa)^2}{4\pi(\delta+\kappa)^2\left(\frac{\kappa^2-\delta^2}{\beta}+\gamma-\alpha\right)}\, ,
\end{align}
where we used the results~\cite{You_2020}
\begin{align}
&\mathcal{A}_{A}^{1,*}=2\sqrt{\frac{\kappa^2-\delta^2}{\beta}+\gamma-\alpha}\,,
\\
&\mathcal{A}_{B}^{1,*}=\frac{\kappa+\delta}{\beta}\mathcal{A}_{A}^{1,*}\,.
\end{align}
We find that the approximate analytical result \red{Eq.~\eqref{theta_analytical}} is in good agreement with the numerical data [see Fig.~\ref{fig:delta_theta}], which confirms the validity of the approximation in the shown parameter regime.

\section{Stochastic dynamics and entropy production for $\phi^*=0$}\label{sec:eprhom}
Here we show how  to compute $\mathcal{S}$ in the demixed phase, i.e. for $\phi^*=0$ up to leading order in $\epsilon$ using
Eq.~\eqref{eq:EPRmodes}. 
As described in Sec.~\ref{sec:homentropy}, the contribution from each mode ${\phi}^j$ can be obtained by computing the covariance matrix of the leading order expansion coefficient of the small noise expansion of ${\psi}^j$, which is defined as 
\begin{align}
\mathrm{Cov(\psi)}_{ij}^k\equiv\Big\langle\Big({\psi}^k_{i}{\psi}^{-k}_j-\langle{\psi}^k_{i}\rangle\langle{\psi}^{-k}_j\rangle\Big)^2\Big\rangle \, .
\end{align}
The expansion coefficients each satisfy an $\epsilon$-independent dynamical equation given by
\begin{align}\label{equ:eomm_a}
\dot{{{\psi}}}^k(t)= \mathbb{A}_k{{\psi}}^k(t)+{\xi}^k\,,
\end{align}
with
\begin{align}
 \mathbb{A}_k=-q_k^2
\begin{pmatrix}
\alpha+\gamma q_k^2  & \kappa-\delta\\
\kappa+\delta & \beta\\
\end{pmatrix} \,,
\end{align}
and 
\begin{align}
\langle\xi^k \xi^l\rangle=2\epsilon \frac{q_l^2}{L}\delta_{k,-l}\,.
\end{align}
Since $\phi^*$ is a linearly stable fixed point, $\mathbb{A}^k$ has full rank for each $k$. Hence, taking the average of Eq.~\eqref{equ:eomm_a} we find $\langle{\psi}^k_i\rangle=0$ for $i\in\{A,B\}$.
Now applying the It\^o formula to ${\psi}^k_{i}{\psi}^{-k}_j$ and averaging, we obtain the 
 so-called Lyaponov equation assigned to Eq.~\eqref{equ:eomm_a} which reads
\begin{align}
    (\mathrm{Cov}(\psi)^{-k}(\bar{\mathbb{A}}_k)^T)_{ij}+(\mathbb{A}_k\mathrm{Cov}(\psi)^k)_{ij}=2\frac{\epsilon}{L}\delta_{ij} \, .
\end{align}
The solutions for the relevant entries read 
\begin{widetext}
\begin{align}
   \left\langle \mathrm{Re}{\psi}^k_A{\psi}^{-k}_B\right\rangle &=\frac{1}{2}\left(\mathrm{Cov}(\psi)^{k}_{AB}+\mathrm{Cov}(\psi)^{-k}_{AB}\right)=-\frac{\kappa(\alpha + \gamma q_k^2+\beta)+\delta[\beta-(\alpha + \gamma q_k^2)]}{[(\alpha + \gamma q_k^2)+\beta)(\delta^2-\kappa^2+(\alpha + \gamma q_k^2)\beta]}\frac{\epsilon}{L}\,,\\
    \left\langle\vert {\psi}^k_A\vert^2\right\rangle&=\mathrm{Cov}(\psi)^{k}_{AA}=\frac{\delta-\kappa}{(\alpha + \gamma q_k^2)}\left\langle \mathrm{Re}{\psi}^k_A{\phi}^{-k}_B\right\rangle+\frac{1}{(\alpha + \gamma q_k^2)}\frac{\epsilon}{L}\,,\\
    \left\langle\vert {\phi}_B^k\vert^2\right\rangle&=\mathrm{Cov}(\psi)^{k}_{BB}=-\frac{\delta+\kappa}{\beta}\left\langle \mathrm{Re}{\psi}^k_A{\psi}^{-k}_B\right\rangle+\frac{1}{\beta}\frac{\epsilon}{L} 
    \, .
\end{align}
\end{widetext}
Noting that $\left\langle{\phi}^{-k}_i{\phi}^k_j\right\rangle=\epsilon\left\langle{\psi}_i^{-k}{\psi}_j^k\right\rangle+\mathcal{O}(\epsilon^2)$, $i,j\in \{A,B\}$ and plugging the result into \red{Eq.~\eqref{eq:EPRmodes}}, we obtain
\begin{align}
   \sigma^k
    =\frac{8\delta^2 \,q_k^2}{\alpha+\beta+\gamma q_k^2}+\mathcal{O}(\epsilon)
    \,.
\end{align}
We note that 
\begin{align}\label{Ajinf}
 \mathbb{A}_k\sim
\begin{pmatrix}
\gamma q_k^2  & \kappa-\delta\\
\kappa+\delta & \beta\\
\end{pmatrix}\,, ~\text{as}~k\rightarrow\infty\,,
\end{align}
which means that fluctuations become effectively decoupled for large wavenumbers. Consequently, their statistics becomes independent of $\alpha$ and of the underlying zero noise solution $\phi^*$. Repeating the above calculation, we find
\begin{align}
    \left\langle \mathrm{Re}{\phi}^k_A{\phi}^{-k}_B\right\rangle&\sim\frac{1}{\gamma q_k^2}\frac{\delta-\kappa}{\beta}\frac{\epsilon}{L}\,,\\
    \left\langle\vert {\phi}^k_A\vert^2\right\rangle&\sim\frac{1}{\gamma q_k^2}\frac{\epsilon}{L}\,,\\
    \left\langle\vert {\phi}_B^k\vert^2\right\rangle&\sim\frac{1}{\beta}\frac{\epsilon}{L}\,,
\end{align}
as $k\rightarrow\infty$, and consequently
\begin{align}
    \sigma^k
    \sim\frac{8\delta^2 \,}{\gamma }+\mathcal{O}(\epsilon)\,, ~\text{as}~k\rightarrow\infty
\end{align}
in the whole parameter range of the phase diagram [as given in Eq.~\eqref{eq:sigmainf}].
Thus, there are always infinitely many, identical, noise-independent contributions with $\sim \delta^2$ in all phases. Physically, however, they have little significance because they come from length scales at which hydrodynamic theory loses its validity, motivating a UV cutoff.

\section{UV cutoff}
\label{Sec:cutoff}
\noindent
Since throughout the phase diagram $\sigma^k\sim \delta^2/\gamma$ as $k\rightarrow\infty$ (see App.~\ref{sec:eprhom}), $\mathcal{S}$ given in  Eq.~\eqref{equ:esum} as a sum over infinitely many such mode contributions, formally takes the value infinity and is thus not well-defined measure of TRSB. 
A way of regularizing $\mathcal{S}$ is by truncating the sum in Eq.~\eqref{equ:esum} at a finite at a wavenumber $k_{\text{\text{max}}}$, 
\begin{align}\label{equ:esum_2}
\mathcal{S}=\sum_{k=1}^{k_{\text{max}}} \sigma^k\, .
\end{align}
Importantly, this method of regularizing does not spoil the genuinely probabilistic interpretation of $\mathcal{S}$ in terms of path probabilities. 

To show this, we first note that Eq.~\eqref{Ajinf} implies that above a sufficiently large wavenumber $k_{\text{max}}$, the dynamics of each mode ${\phi}^k$ evolves independently of that of all other modes.
Hence, the path probability decomposes approximately into the following product
\begin{align}
  &  \mathbb{P}\left[\big\{\{{\phi}^k\}_{t\in [0,T]}\big\}_{k>0}\right]
    \nonumber \\
  & ~ \approx\mathbb{P}\left[\big\{\{{\phi}^k\}_{t\in [0,T]}\big\}_{k_{\text{max}}\geq  k>0}\right]\prod\limits_{k>k_{\text{max}}}\mathbb{P}\left[\{{\phi}^k\}_{t\in [0,T]}\right]\,.
\end{align}
Consequently, the series representation of $\mathcal{S}$ in Eq.~\eqref{equ:esum_2} can be approximately decomposed as
\begin{align}
\mathcal{S}
\approx\mathcal{S}\left[\big\{\{{\phi}^k\}_{t\in [0,T]}\big\}_{k_{\text{max}}\geq  k>0}\right]+\sum\limits_{k>k_{\text{max}}} \mathcal{S}\left[\{{\phi}^k\}_{t\in [0,T]}\right]
\end{align}
with
\begin{align}
\mathcal{S}\left[\big\{\{{\phi}^k\}_{t\in [0,T]}\big\}_{k_{\text{max}}\geq  k>0}\right]\approx\sum\limits_{k=1}^{k_{\text{max}}} \sigma^k\,,
\end{align}
and
\begin{align}
\mathcal{S}\left[\{{\phi}^k\}_{t\in [0,T]}\right]\approx \sigma^k~~\text{for}~k>k_{\text{max}}\,.
\end{align}
This allows us to conclude that $\mathcal{S}$ in Eq.~\eqref{equ:esum_2} can be understood as the sum of the entropy productions of an infinite number of individual processes, each of which is finite and is therefore naturally divergent. 

Since the spurious high modes with $\sigma^k\sim \delta^2/\gamma$ as $k\rightarrow\infty$, represent physically irrelevant or even unphysical degrees of freedom that all produce entropy, it is actually rather the regularized, truncated series which should be understood as a realistic measure of TRSB (in the sense of a path probability ratio) of the physically pertinent mesoscale dynamics.

\bibliography{bib}

\end{document}